\documentclass[a4paper,10pt,aps,pre,twocolumn,superscriptaddress,showpacs,amsmath,amssymb]{revtex4}
\usepackage{graphicx,color}

\newcommand{\ahum}[1]{``#1''}
\newcommand{\eq}[1]{Eq.~(\ref{#1})}
\newcommand{\fig}[1]{Fig.~\ref{#1}}
\newcommand{\sect}[1]{Section~\ref{#1}}
\newcommand{\olcite}[1]{Ref.~\onlinecite{#1}}

\newcommand{\ie}{i.e.,~}
\newcommand{\dd}{\mathrm{d}}
\newcommand{\pd}[2]{\frac{\partial #1}{\partial #2}}

\newcommand{\mean}[1]{\langle #1 \rangle}
\newcommand{\Int}[1]{\int\dd #1\;}
\newcommand{\IInt}[3]{\int_{#2}^{#3}\dd #1\;}
\newcommand{\Path}[1]{\int[\dd#1]\;}

\newcommand{\mat}[1]{\mathsf #1}

\newcommand{\al}{\alpha}
\newcommand{\gam}{\gamma}
\newcommand{\eps}{\varepsilon}
\newcommand{\kap}{\kappa}
\newcommand{\lam}{\lambda}

\newcommand{\sig}{\sigma}
\newcommand{\ra}{\rightarrow}

\newcommand{\im}{\text{i}}
\newcommand{\eb}{\eps_\text{b}}
\newcommand{\ebx}{\eps_\text{b}^\text{cx}}
\newcommand{\eA}{\epsilon_\text{A}}
\newcommand{\eB}{\epsilon_\text{B}}
\newcommand{\x}{\mathbf r}
\newcommand{\q}{\mathbf q}
\newcommand{\Q}{\mathcal Q}
\newcommand{\Ha}{\mathcal H}

\newcommand{\rgh}{\xi_\perp}     
\newcommand{\cor}{\xi_\parallel} 
\newcommand{\id}{\mathbf 1}

\newcommand{\hp}{h_\text{P}}

\begin{document}

\title{Random pinning limits the size of membrane adhesion domains}

\author{Thomas Speck}
\affiliation{Institut f\"ur Theoretische Physik II: Weiche Materie,
  Heinrich-Heine-Universit\"at D\"usseldorf, Universit\"atsstra\ss e 1,
  D-40225 D\"usseldorf, Germany}
\author{Richard L. C. Vink}
\affiliation{Institute of Theoretical Physics, Georg-August-Universit\"at 
  G\"ottingen, Friedrich-Hund-Platz~1, D-37077 G\"ottingen, Germany}


\begin{abstract}
  Theoretical models describing specific adhesion of membranes predict (for
  certain parameters) a macroscopic phase separation of bonds into adhesion
  domains. We show that this behavior is fundamentally altered if the membrane
  is pinned randomly due to, e.g., proteins that anchor the membrane to the
  cytoskeleton. Perturbations which locally restrict membrane height
  fluctuations induce quenched disorder of the random-field type. This
  rigorously prevents the formation of macroscopic adhesion domains following
  the Imry-Ma argument [Y.~Imry and S.~K.~Ma, Phys.~Rev.~Lett. {\bf 35}, 1399
  (1975)]. Our prediction of random-field disorder follows from analytical
  calculations, and is strikingly confirmed in large-scale Monte Carlo
  simulations. These simulations are based on an efficient composite Monte
  Carlo move, whereby membrane height and bond degrees of freedom are updated
  simultaneously in a single move. The application of this move should prove
  rewarding for other systems also.
\end{abstract}



\maketitle

\section{Introduction}

The fate of living cells is regulated through interactions with other cells
and with the extracellular matrix (ECM). Through receptor-ligand bonds formed
by specific proteins the cell adheres to the ECM forming adhesion domains
(clusters of closed bonds). Not all adhesion domains are focal adhesions but
these are particularly well studied and relevant. Focal adhesions are involved
in the transmission of signals and mechanical forces, and play key roles in
cell anchorage and migration~\cite{geig01,pars10}. Consequently, understanding
how adhesion domains form, and the factors that control their size, shape, and
growth~\cite{nico04,gov06}, is of profound practical importance. Extensive
studies have been performed on theoretical models for single bond
dynamics~\cite{bell78}, collective dynamics of discrete
bonds~\cite{weik02,krob07,spec10b,weil10,fara11}, and employing effective
potentials~\cite{lipo96,zhan08,atil09}; as well as experimentally on
cell-mimetic model systems~\cite{tana05,moss07} such as lipid bilayer vesicles
with embedded ligands brought in the vicinity of receptors tethered to
supported membranes~\cite{brui00,cuve04,reis08,limo09,smit09}.

\begin{figure}[b!] 
  \centering 
  \includegraphics[width=\columnwidth]{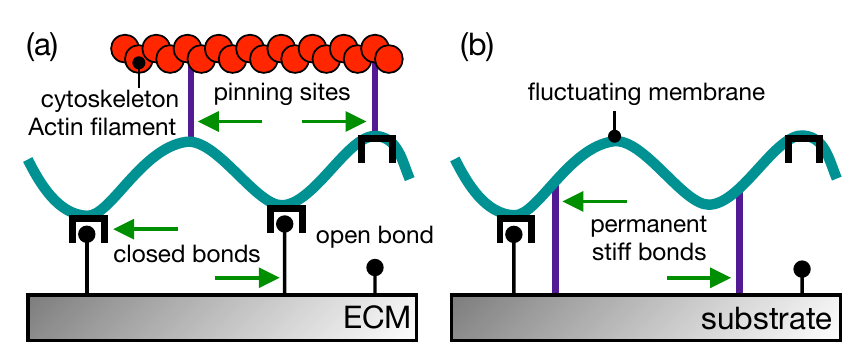} 
  \caption{\label{fig:mem} (Color online) Sketch of two typical situations in
    which the formation of macroscopic adhesion domains is prevented due to
    the presence of random fields: (a)~A membrane of a biological cell adheres
    to the ECM via receptor-ligand bonds, which may dynamically open and
    close. The membrane is freely fluctuating except at the anchoring sites
    (pinning sites) attaching the membrane to the cytoskeleton. (b)~An
    analogous \emph{in vitro} example in which the pinning sites are formed by
    stiff and permanently closed bonds (tethers) between membrane and 
substrate.}
\end{figure}

While theoretical models predict a macroscopic phase separation, adhesion 
domains in cells are typically finite in size. It has been proposed that 
trapping in membrane corrals (or compartments) reduces the mobility of 
receptors~\cite{kusu05}, which becomes a limiting factor for the growth of 
adhesion domains. The purpose of this paper is to provide an alternative 
explanation based on the observation that, due to the interactions with the ECM, 
membrane height fluctuations are locally suppressed. These \ahum{pinning sites} 
induce quenched disorder of the \emph{random-field} type which prevents 
macroscopic domain formation in $d=2$ dimensions~\cite{imry75, imbrie:1984, 
bricmont.kupiainen:1987, aizenman.wehr:1989}. Hence, based on the fundamentals 
of statistical physics alone, adhesion domains of finite size are implied in a 
wide variety of contexts. The mere presence of random-field disorder provides a 
robust mechanism controlling the size of adhesion domains, irrespective of the 
details of the many and complex molecular processes between cell and ECM.

The generic situation that we envision is sketched in~\fig{fig:mem}(a), which 
shows a fluctuating membrane adhering to a substrate via receptor-ligand bonds. 
The crucial ingredients are the pinning sites at which the membrane height is 
assumed to be fixed. In a biological cell, such pinning sites correspond, e.g., 
to the locations where the cytoskeleton anchors to the membrane. For an 
experimental verification of our predictions \emph{in vitro} we propose a 
bilayer adhered to a substrate or supported membrane with a (small) fraction of 
the receptor-ligand bonds permanently closed and of high stiffness 
[\fig{fig:mem}(b)]. We will show that in situations resembling those 
of~\fig{fig:mem} adhesion domains remain finite in size for any finite 
concentration of pinning sites, provided the spatial distribution of pinning 
sites is random.

\section{Model}

To show how the random-field disorder in the specific adhesion of membranes
comes about we consider a class of simple models as reviewed in
\olcite{weik07}. These coarse-grained models contain the minimal ingredients
that we require to address the influence of membrane pinning on the statistics
of adhesion domain formation. In particular, we assume the membrane to be a
two-dimensional sheet characterized by a bending rigidity, and to be in
thermal equilibrium with its surroundings. The ligand-receptor bonds and
pinning sites are treated as point particles. We neglect all active processes
and, in the case of focal adhesions, the influence that stresses applied to
the substrate may have on the development of adhesion domains~\cite{nico04}.

In the Monge representation a membrane patch with projected area $A=L^2$ is
described through its separation profile $h(\x)$ describing the height of the
membrane at position $\x=(x,y)$, $0 \leqslant x,y < L$, measured with respect
to some (arbitrarily chosen) reference height. The effective Hamiltonian reads
\begin{equation}
  \label{eq:H}
  \Ha_0[h(\x)] = \IInt{^2\x}{A}{} \left\{ \frac{\kap}{2}[\nabla^2 h(\x)]^2
    + \frac{\gamma}{2}[h(\x)]^2 \right\}.
\end{equation}
The first term governs the bending energy, with $\kap$ the membrane bending
rigidity, which we assume is the dominant contribution to the Helfrich
energy~\cite{helf78}. The second term is the lowest-order expansion of the
non-specific interactions between membrane and substrate whose strength is
denoted $\gamma$. In our treatment, the minimum of the non-specific potential
is thus taken to be the reference height from which $h(\x)$ is measured. These
non-specific interactions arise due to volume exclusion, van der Waals forces,
and the possible formation of an electrostatic double layer, as well as an
effective pressure due to the restricted volume the membrane can move in. 

The parameters $\kap$ and $\gamma$ define a length, $\cor \equiv 
(\kap/\gamma)^{1/4}$, which sets the scale over which the membrane height 
fluctuations are correlated~\cite{spec10b},
\begin{equation}
  \label{eq:m}
  m(|\x-\x'|) \equiv \mean{h(\x)h(\x')}_0 \approx \rgh^2u(|\x-\x'|/\cor),
\end{equation}
see appendix~\ref{sec:corr}. The scaling function is defined in terms of the 
Kelvin function, $u(x) = -(4/\pi) \, {\rm kei}_0(x)$; it obeys $u(0)=1$ and 
decays to zero exponentially fast. The amplitude of the correlations is set by 
the thermal roughness $\rgh\equiv\cor/\sqrt{8\beta\kap}$, 
$\beta\equiv(k_\text{B}T)^{-1}$, with temperature $T$ and Boltzmann constant 
$k_\text{B}$. The brackets $\mean{\cdots}_0$ denote the thermal average with 
respect to the Hamiltonian Eq.~\eqref{eq:H}, \ie in the absence of 
ligand-receptor bonds and pinning sites.

Now imagine a set of $N$ receptor-ligand pairs at positions $R\equiv\{\x_1,
\x_2, \ldots, \x_N \}$ embedded in the membrane and substrate that can form
bonds. The arguably simplest model is to assign a linear energy to closed
bonds with stiffness~$\al$~\cite{weik02}. The total Hamiltonian then reads
\begin{equation}
  \label{eq:HN}
  \Ha_N=\Ha_0+\sum_{i=1}^N b_i[\al h(\x_i)-\eb],
\end{equation}
where $b_i=0$ if the $i$th bond is open, and $b_i=1$ if the bond is
closed. The parameter $\eb=\eb^0-\al d$ is the shifted binding energy the
system gains through forming a bond, where $\eb^0$ is the bare binding energy
and $d$ is the separation between substrate and the minimum of the
non-specific potential. In addition to the receptor-ligand pairs, we assume
the presence of $n$ pinning sites at positions
$P\equiv\{\x_{N+1},\ldots,\x_M\}$ where the membrane height is fixed to
$h=\hp$. We take the two sets $R$ and $P$ to be disjoint with their union thus
holding $M=N+n$ distinct sites.

\section{Theoretical mapping}

We now calculate the free energy as a function of the bond variables $\{b_i\}$
through integrating out the height fluctuations under the constraints
$h(\x_i)=h_i$ for sites $\x_i\in R$ and $h(\x_i)=\hp$ for $\x_i\in P$. We
follow the standard procedure and implement these $M$ constraints through
$\delta$-functions~\cite{li91}; for the detailed calculation see
appendix~\ref{sec:free}.

For clarity of the presentation, in the following we consider the case
$\hp=0$. The free energy then reads
\begin{multline}
  \label{eq:G:int}
  G(\{b_i\}; R\cup P) = -\beta^{-1}\ln\Int{h_1\cdots\dd h_N} \times \\
  \exp\left\{-\frac{1}{2} \sum_{i,j=1}^N (\mat m^{-1})_{ij} h_i h_j 
    - \beta\sum_{i=1}^N b_i[\al h_i-\eb] \right\}.
\end{multline}
The dependence on the positions $R$ and $P$ is encoded in the matrix $\mat m$,
whose components are given by $m_{ij} \equiv m(|\x_i-\x_j|)$. Performing the
final integrations we obtain
\begin{equation}
  \label{eq:G}
  G = -\sum_{i=1}^N \sum_{j=i+1}^N \eps_{ij} b_i b_j - 
  \sum_{i=1}^N \mu_i b_i, \quad \eps_{ij} \equiv \beta\al^2 \hat m_{ij},
\end{equation}
with $\mu_i\equiv\eb+\beta\al^2\hat m_{ii}/2$, and where we have introduced a 
new matrix $\hat{\mat m}$. While Eq.~\eqref{eq:G:int} contains the full matrix 
$\mat m$, the sums run only over the first $N$ sites corresponding to the bonds 
and excluding the pinning sites. The matrix $\hat{\mat m}$ is obtained by 
inverting the submatrix formed by the first $N$ rows and $N$ columns of $\mat 
m^{-1}$. Note also that we have dropped an additional term in \eq{eq:G} which 
depends on the geometry of both the bonds and the pinning sites but not on the 
state $\{b_i\}$ of the bonds.

The key result here is that the free energy in \eq{eq:G} is isomorphic to the
Ising lattice gas with couplings $\eps_{ij}\propto\al^2$ and a
\emph{site-dependent} effective chemical potential $\mu_i$ that also depends
on the stiffness $\al$. There are two effects due to membrane undulations:
First, single bond formation is assisted ($\mu_i>\eb$) since the system can
access configurations with lower energy. Second, bonds couple in a manner that
enhances clustering: a closed bond pulls down the membrane locally making it
easier for nearby bonds to also close. In such bound patches (adhesion
domains) membrane fluctuations are hindered and, therefore, the entropy is
decreased. The phase behavior of the system is thus determined by the
competition between this loss of entropy, the mixing entropy of bonds, and the
gain in binding energy. For small $\cor$ only nearest neighbors interact
directly; by increasing $\cor$ (\ie for stiffer membranes) the thermal
roughness $\rgh$ determining the strength of the coupling is diminished but
more and more bonds become coupled.


\subsection{Membrane without pinning}

The result Eq.~\eqref{eq:G} holds for any geometry of bonds and pinning sites.
In the absence of pinning sites ($n=0$) clearly $\hat{\mat m}=\mat m$ and the
chemical potential becomes spatially uniform, $\mu_i=\eb+\eps_0/2\equiv\mu_0$,
where
\begin{equation}
  \label{eq:eps}
  \eps_0 \equiv \beta(\al\rgh)^2 = \al^2/(8\sqrt{\kap\gam}) 
\end{equation}
is the effective coupling energy due to the membrane undulations. We now
specialize to the situation where the positions $R$ of receptor-ligand pairs
form a regular square lattice with lattice spacing $a\sim\cor$. The phase
behavior in this case is known~\cite{lipo96,weik02}: For sufficiently high
$\al>\al^\ast$ there is a first order phase transition from a bound state
($\phi\sim1$) with a high density of closed bonds
\begin{equation}
  \label{eq:phi}
  \phi \equiv \frac{1}{N} \sum_{i=1}^N b_i \quad,
\end{equation}
to an unbound state with a low density of closed bonds
($\phi\sim0$). Precisely at $\al=\al^\ast$ the system shows critical behavior
which, by virtue of the mapping of \eq{eq:G}, belongs to the universality
class of the $d=2$ Ising model.


\subsection{The pinned membrane}

\begin{figure}[t]
  \centering
  \includegraphics{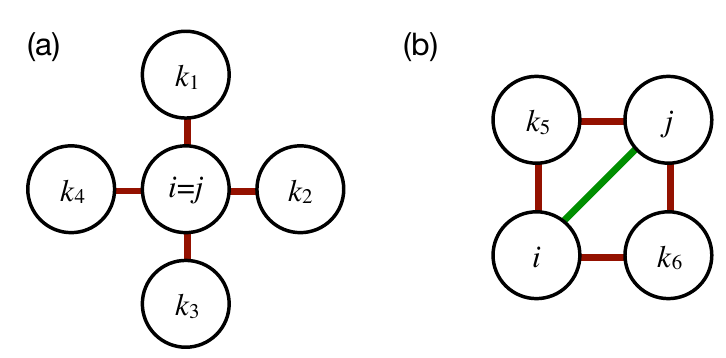}
  \caption{\label{fig:count} (Color online) Graphical aid to identify the
    non-vanishing components of the matrices $\widehat{{\mat A}^2}$ and
    ${\hat{\mat A}}^2$ appearing in \eq{eq:1} with bond sites $i$ and $j$:
    (a)~For $i=j$ the sites $k_1,\dots,k_4$ that are nearest neighbors of $i$
    are counted. (b)~For the case that $i$ and $j$ are next-nearest neighbors
    the sites $k_5$ and $k_6$ that are common nearest neighbors of both $i$
    and $j$ are counted.}
\end{figure}

We now come to the main result of this paper, where the fate of the mapping
of \eq{eq:G} in the presence of pinning sites $(n>0)$ is considered. In this
case, we must distinguish between the set $R$ of sites where receptor-ligand
bonds may form, and the set $P$ of sites at which the membrane is pinned. We
furthermore restrict our calculation of $\hat{\mat m}$ to nearest and
next-nearest neighbor interactions. We first write $\mat
m=\rgh^2[\id+\eA\mat A+\eB\mat B]$, where the $M \times M$ matrices $\mat A$
and $\mat B$ correspond to nearest and next-nearest neighbor interactions,
respectively. The components $A_{ij}$ ($B_{ij}$) equal one if the two sites
$i$ and $j$ are nearest (next-nearest) neighbors, and zero otherwise. The
coefficients are set by the value of the scaling function at the nearest and
next-nearest neighbor distance, $\eA\equiv u(a/\cor)$ and $\eB\equiv u(\sqrt
2a/\cor)$, respectively. We expand the inverse as
\begin{equation}\label{eq:app}
  (\mat m/\rgh^2)^{-1} \approx 
  \id - \eA\mat A - \eB \mat B + \eA^2 \mat A^2, 
\end{equation}
where all terms with coefficients $<\eA^2$ have been dropped. We now replace,
in \eq{eq:app}, the $M \times M$ matrices $\{\id, \mat A, \mat B\}$ by the $N
\times N$ sub-matrices $\{\hat{\id}, \hat{\mat A}, \hat{\mat B}\}$ obtained by
discarding the $n$ upper rows and columns corresponding to pinning sites;
taking the inverse of the resulting matrix yields the desired matrix
$\hat{\mat m}$ that appears in the mapping of \eq{eq:G}
\begin{equation}\label{eq:1}
 \hat{\mat m} / \rgh^2 \approx
 \hat{\id} + \eA \hat{\mat A} + \eB \hat{\mat B} +
 \eA^2 \left( {\hat{\mat A}}^2 - \widehat{{\mat A}^2} \right),
\end{equation}
where all higher order terms have again been dropped. 

Due to the pinning sites, the last term in \eq{eq:1} does \emph{not} vanish 
since the \ahum{hat} and \ahum{square} operations do not commute. First we 
evaluate $(\widehat{{\mat A}^2})_{ij} = \sum_{k=1}^M A_{ik}A_{kj}$ where, in the 
summation, only two terms survive (\fig{fig:count}). The first term corresponds 
to $i=j$; a non-zero contribution implies that $k$ needs to be a nearest 
neighbor of this site. On a square lattice there are four such sites labeled 
$k_1,\dots,k_4$ in \fig{fig:count}(a). The second term arises when $i$ and $j$ 
are distinct. The only combination with a non-zero contribution is when $i$ and 
$j$ are both nearest-neighbor of the same site $k$, which implies that $i$ and 
$j$ are next-nearest neighbors. Two such sites can be identified, labeled $k_5$ 
and $k_6$ in \fig{fig:count}(b). Hence,
\begin{equation}
  (\widehat{{\mat A}^2})_{ij} = 4\delta_{ij} + 2 B_{ij},  
\end{equation}
with $\delta_{ij}$ the Kronecker symbol. The components of the square of the
reduced nearest-neighbor matrix are $({\hat{\mat A}}^2)_{ij} = \sum_{k=1}^N
A_{ik}A_{kj}$, where the sum over $k$ now excludes the pinning sites. We can
identify the non-vanishing terms as in \fig{fig:count} provided we ignore the
sites $k_i$ that are pinned. We thus obtain
\begin{equation}
  ({\hat{\mat A}}^2)_{ij} = (4-\eta_i) \delta_{ij} + \chi_{ij} B_{ij}  
\end{equation}
with \emph{stochastic} variables $\eta_i$ and $\chi_{ij}$ set by the local
environment of pinning sites. In particular, $\eta_i$ is the number of
nearest-neighbors of site $i$ that are pinned; the possible values of
$\chi_{ij}=0,1,2$ correspond to, respectively, the case where both $k_5$ and
$k_6$ are pinned, only one of those sites is pinned, and neither one of them
being pinned.

We now have all the ingredients needed to discuss the mapping of \eq{eq:G} in
the presence of pinning sites. The first observation is that the
nearest-neighbor coupling is \emph{not} affected. For sites $i$ and $j$ that
are nearest-neighbors, $\eps_{ij}=\eps_0\eA$ as before, by virtue of
\eq{eq:1}. In contrast, the chemical potential is affected, and now depends on
the local environment via $\eta_i$
\begin{equation}\label{eq:pp}
  \mu_i = \mu_0 - \eta_i\Delta, \qquad
  \Delta \equiv \eps_0\eA^2 / 2, 
\end{equation}
where $\mu_0$ is the effective chemical potential in the absence of pinning
sites. Provided the pinning sites are immobile and randomly distributed,
\eq{eq:pp} corresponds to a quenched random-field. The effective chemical
potential is thus reduced, implying that the closing of bonds has become more
difficult. The physical picture is that, since the membrane is pinned to the
minimum of the non-specific potential $(\hp=0)$, closed bonds cannot
\ahum{pull down} the membrane as easily as before. This impedes the above
mentioned facilitation of bond formation due to undulations, and consequently
the effective chemical potential is reduced. Choosing a sufficiently negative
$\hp<0$, the opposite situation of an increased local chemical potential may
also be realized.

We typically consider a low pinning density $\rho \equiv n/N \ll 1$, such that
$\eta_i$ effectively becomes a binary random variable with values
$\eta_i=0,1$. In this limit, the average chemical potential is $[\mu_i] =
\mu_0 - \rho\Delta$, where $[\cdot]$ denotes a disorder average (\ie an
average over many different samples of pinning sites). The random-field
strength is set by the disorder fluctuations $\sqrt{[\mu_i^2] - [\mu_i]^2}
\approx \sqrt{\rho}\Delta$, which thus is weak compared to the
nearest-neighbor coupling. Nevertheless, given that an infinitesimally weak
random-field is sufficient to destroy macroscopic domain formation in two
dimensions~\cite{imry75, imbrie:1984, bricmont.kupiainen:1987,
  aizenman.wehr:1989}, we expect that even a low pinning density will
drastically affect adhesion domain formation. Finally, note that in addition
to the dominant random-field disorder, the pinning sites also induce a
marginal perturbation. For next-nearest neighbors $i$ and $j$, the coupling
$\eps_{ij}=\eps_0[\eB+\eA^2(\chi_{ij}-2)]$ also becomes a random variable
corresponding to \emph{random-bond} disorder, which does not destroy
macroscopic domain formation. Possible interactions between receptor-ligand
bonds, e.g., due to size mismatch, will change the couplings $\eps_{ij}$ but
not the fact that a random-field is induced.

\section{Simulations}

\subsection{Model and methods}

\subsubsection{Discretized membrane model with pinning sites}
\label{sim:qd_howto}

We now perform Monte Carlo (MC) simulations using a discretized version of our 
model Hamiltonian. We mostly simulate on periodic $N=L \times L$ lattices, and 
to each lattice site $i$ we assign a real number $h_i$ to denote the local 
membrane height, and a bond variable $b_i$ to denote whether the bond at the 
site is open ($b_i=0$) or closed ($b_i=1$). The Hamiltonian may then be written 
as
\begin{equation}
\label{eq:sim}
 {\cal H}_N = \sum_i \left[ \frac{\kappa}{2} 
 (\nabla^2 h_i)^2 + \frac{\gamma}{2} h^2_i 
 + b_i (\al h_i - \eb) \right],
\end{equation}
where the sum extents over all lattice sites. To compute the Laplacian, we use
the finite-difference expression
\begin{equation}\label{eq:fin}
 \nabla^2 h_i \equiv \lam_i = -4h_i + \sum_{j \in {\rm nn}(i)} h_j,
\end{equation}
where the sum in the last term is over the $(z=4)$ nearest neighboring (nn)
sites of site $i$. In this section the inverse temperature $\beta$ is absorbed
into the coupling constants $\{\kappa,\gamma,\al,\eb\}$ of \eq{eq:sim}, while
the lattice constant will be the unit of length.

To incorporate pinning, a fraction $\rho$ of randomly selected sites is placed
in the set $P$ of pinning sites. The sites $i \in P$ have their height
variable set to $h_i=\hp$ at the start of the simulation, and these heights
are not permitted to change during the course of the simulation (\ie MC moves
that change the membrane height are not applied to the pinning sites). In
contrast to the theoretical derivation we do allow for bonds to open and close
at the pinning sites. In the simulations, the sets $R$ and $P$ therefore
overlap. This does not affect the phase behavior of \eq{eq:sim} but makes the
data analysis easier since the available area for bonds then always equals
$L^2$ as opposed to $(1-\rho)L^2$.

\subsubsection{Composite Monte Carlo move}
\label{sim:the_moves}

The \ahum{standard approach} to simulate \eq{eq:sim} is to use MC moves that 
either (i) propose a small random change to a randomly selected height variable, 
or (ii) change the state of a randomly selected bond variable, and to accept 
these changes with the Metropolis criterion~\cite{weik02}. This approach is not 
efficient because the height and bond degrees of freedom are correlated: at 
sites containing a closed bond, the membrane height will be lower, and {\it vice 
versa}. Hence, after a proposed change in $b_i$, the corresponding height $h_i$ 
is likely to be energetically unfavorable, and so move (ii) has a high chance of 
being rejected.

To circumvent this problem, we use a composite MC move whereby the bond and 
height degrees of freedom are changed simultaneously. The key observation is 
that \eq{eq:sim} is quadratic in the height variables. For a given site $i$, 
imagine to replace the corresponding membrane height by $h_i \mapsto h_i' = h_i 
+ \delta$. The energy as function of $\delta$ is quadratic, $E(\delta) = 
\Lambda_2 \delta^2 + \Lambda_1 \delta + \rm constant$, with coefficients given 
by
\begin{gather}
 \Lambda_2 = 10 \kappa + \gamma/2, \\
 \Lambda_1 = \gamma h_i +  \alpha b_i - \kappa \left( 
 4\lam_i -  \sum_{j \in {\rm nn}(i)} \lam_j \right).
\end{gather}
Hence, there is an optimal deviation, $\delta_{\rm opt}=-\Lambda_1 /2\Lambda_2$, 
at which the energy becomes minimized. In our MC simulations, we exploit this 
property by selecting the membrane height deviations $\delta$ from a Gaussian 
distribution around the optimal value
\begin{equation}\label{eq:draw}
 P(\delta|\delta_{\rm opt}) \propto 
 e^{-(\delta-\delta_{\rm opt})^2 / 2\sigma^2},
\end{equation}
where standard deviation $\sigma^2=1/2\Lambda_2$ is used (other choices for 
$\sigma^2$ are valid also, but we believe this one is optimal as it closely 
matches the thermal fluctuations).

The composite MC move that we use to change the state of bond variables proceeds 
as follows:
\begin{enumerate}
\item Randomly select a lattice site $i$, and compute the optimal height 
deviation $\delta_{\rm opt}$.

\item Change the state of the bond variable $b_i \mapsto b_i'$, and compute the 
new optimal height deviation $\delta_{\rm opt}'$. Propose a new height, $h_i 
\mapsto h_i' = h_i + \delta$, with $\delta$ drawn from $P(\delta|\delta_{\rm 
opt}')$ of \eq{eq:draw}.

\item Accept the proposed values $(b_i',h_i')$ with the Metropolis criterion
\begin{equation}\label{eq:metro}
P_{\rm acc} = \min \left[1,
\frac{ P(0|\delta_{\rm opt}) }{ P(\delta|\delta_{\rm opt}') } 
e^{-({\cal H}_N' - {\cal H}_N)} \right],
\end{equation}
with ${\cal H}_N$ the energy of the configuration at the start of the MC move, 
and ${\cal H}_N'$ the energy of the proposed configuration (the ratio of 
Gaussian probabilities is needed to restore detailed balance).
\end{enumerate}
Note that our composite move is ergodic, and thus by itself constitutes a 
valid MC algorithm. Nevertheless, we still found it useful to also implement the 
non-composite variant, whereby the membrane height is updated without changing 
the corresponding bond variable. In terms of the composite move above, this 
corresponds to $\delta_{\rm opt}' = \delta_{\rm opt}$, while in step (2) the 
bond variable $b_i$ is not changed. In what follows, composite to non-composite 
moves are attempted in a ratio 1:2, respectively. 

In the MC moves above, we restrict the selection of the optimal height value 
$h_i$ to the single site~$i$. An obvious generalization is to also optimally 
select the height values of nearby sites, \ie on a $l \times l$ plaquette around 
site~$i$. The above moves correspond to $l=1$, but we have used the version with 
$l=3$ also; the latter is slightly more efficient in cases where bound and 
unbound membrane patches coexist. Obviously, the case $l>1$ is more complex to 
implement, as it involves minimizing a quadratic form of $l^2$ variables. 
Furthermore, the accept criterion \eq{eq:metro} needs to be modified (the 
prefactor now becomes the product of $l^2$ Gaussian probability ratios).

\subsubsection{Order parameter distribution}

A key output of our simulations is the order parameter distribution $P(\phi)$ 
(OPD), which is defined as the probability to observe a system with a fraction 
of closed bonds~$\phi$, with $\phi$ given by \eq{eq:phi}. We emphasize that 
$P(\phi)$ depends on all the coupling constants that appear in \eq{eq:sim}, as 
well as on the system size $L$. To ensure that $P(\phi)$ is sampled over the 
entire range $0 \leq \phi \leq 1$, we combine our simulations with an umbrella 
sampling scheme~\cite{virnau.muller:2004}. We also use histogram 
reweighting~\cite{ferrenberg.swendsen:1988} in the binding energy: having 
measured $P(\phi)$ for some value $\eb=\eb^0$, we extrapolate to different 
values $\eb=\eb^1$ using the relation
\begin{equation}
\label{eq:hr}
  P(\phi|\eb=\eb^1) \propto P(\phi|\eb=\eb^0) \,
  e^{L^2 \left(\eb^1 - \eb^0\right) \phi}.
\end{equation}
In a similar way we also use histogram reweighting in the coupling constant
$\al$, which is slightly more complex to implement as it requires separate
storage of the fluctuations in $\sum_i b_i h_i$, \ie the third term of
\eq{eq:sim}.

We emphasize that, in the presence of pinning, $P(\phi)$ may also depend on the 
particular sample of pinning sites. For an accurate analysis, it then becomes 
necessary to average simulation results over many different random positions 
(samples) of the pinning sites.

\subsection{Membrane without pinning}
\label{nopin}

We first simulate a membrane without pinning ($\rho=0$) using $\kappa=\gamma=1$ 
in the Hamiltonian of \eq{eq:sim}. This case was considered extensively in 
\olcite{weik02}, the main conclusion being that macroscopic adhesion domains are 
observed for $\al > \al^\ast$. We revisit this case to also determine the 
universality class, as well as the line tension between coexisting domains.

\subsubsection{Critical behavior}

We first determine the critical value $\al^\ast$ via finite-size scaling of the 
order parameter $m = \mean{|\delta\phi|}$, the susceptibility $\chi = L^2 
\left(\mean{\delta\phi^2} - \mean{|\delta\phi|}^2 \right)$, and the Binder 
cumulant $U_4 = \mean{\delta\phi^2}^2 / \mean{\delta\phi}^4$. Here, $\delta\phi 
\equiv \phi - \mean{\phi}$, with thermal averages $\mean{\phi^n} = \int_0^1 
\phi^n P(\phi) d\phi$, where it is assumed that $P(\phi)$ is normalized. We 
emphasize that $m,\chi,U_4$ are to be computed at the coexistence value of the 
binding energy $\eb=\ebx$. An accurate numerical criterion to determine 
the latter is to tune $\eb$ such that the fluctuations in $\phi$ become 
maximized~\cite{citeulike:8864903}
\begin{equation}
  \label{eq:cx}
  \ebx : \mean{\phi^2} - \mean{\phi}^2 \to {\rm max},
\end{equation}
which may conveniently be done using the histogram reweighting formula of 
\eq{eq:hr}. 

\begin{figure}
\centering
\includegraphics[width=\columnwidth]{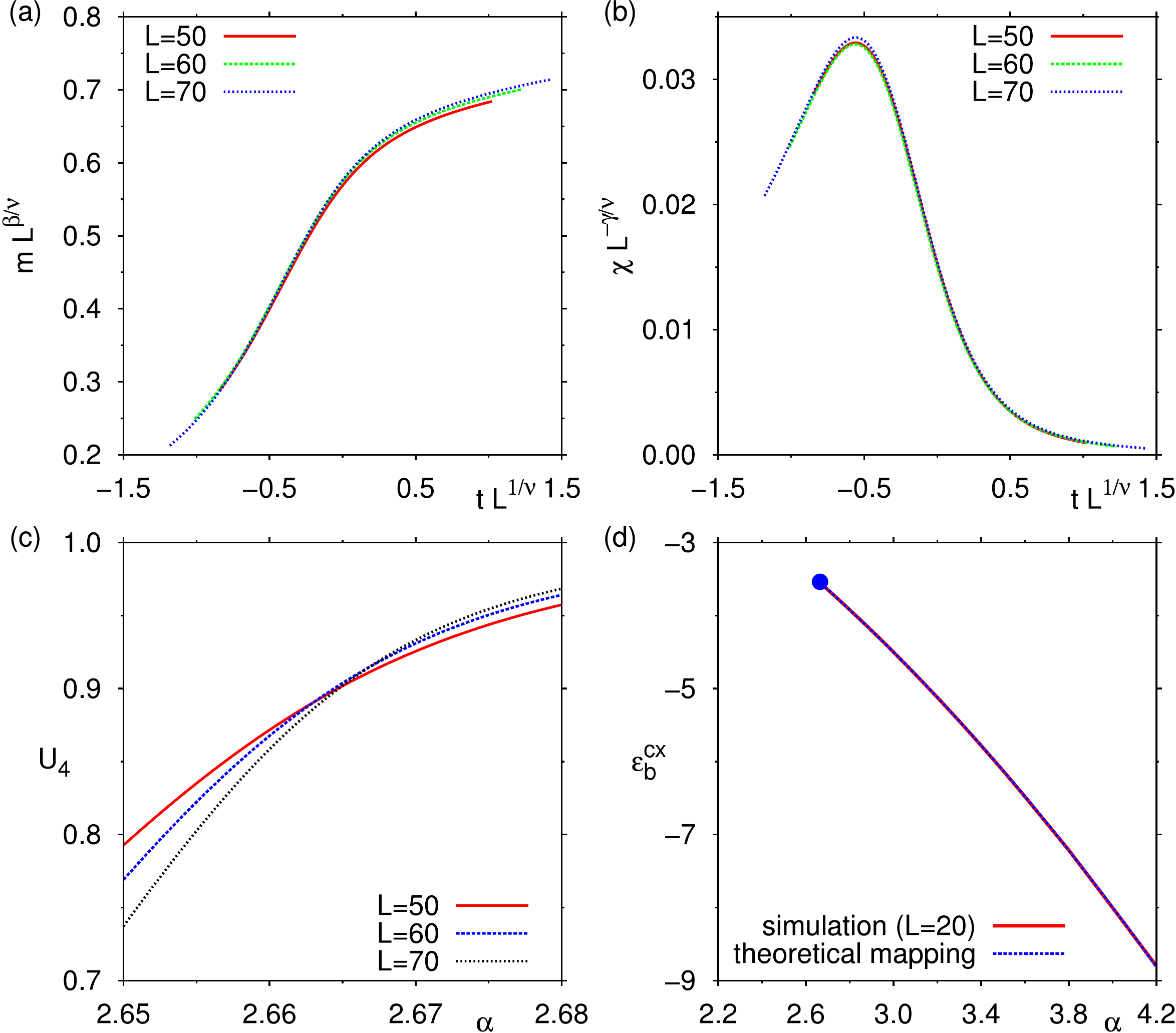}
\caption{\label{fss} (Color online) Finite-size scaling results for the membrane 
without pinning sites. The scaling plots of the order parameter (a) and 
susceptibility (b) confirm 2D Ising universality, as indicated by the collapse 
of the data from different system sizes. The intersection of the Binder cumulant 
curves for different system sizes (c) yields an estimate of $\al^\ast$ 
consistent with the scaling plots. The variation of $\ebx$ with $\al$ 
(d) conforms to the theoretical prediction, where the dot marks the critical 
point.}
\end{figure}

In \fig{fss}(a), we show the scaling plot of the order
parameter~\cite{newman.barkema:1999}, \ie curves of $m L^{\beta/\nu}$ versus
$t L^{1/\nu}$, $t=\al/\al^\ast-1$, using 2D Ising critical exponents
$\beta=1/8$, $\nu=1$, and with $\al^\ast = 2.665$ obtained by tuning until the
curves for different $L$ collapsed (note: we use the \ahum{standard symbol}
$\beta$ for the order parameter critical exponent, which is not to be confused
with the inverse temperature). The fact that the data collapse confirms 2D
Ising universality, and our estimate of $\al^\ast$ is in good agreement with
\olcite{weik02}. In \fig{fss}(b), we show the corresponding scaling plot of
the susceptibility, using the 2D Ising value $\gamma=7/4$, while for
$\al^\ast$ the above estimate was used. A data collapse is again observed,
providing further confirmation of 2D Ising universality. In \fig{fss}(c), we
plot the Binder cumulant $U_4$ versus $\al$. In agreement with a critical
point, the curves for different $L$ intersect at $\al^\ast$.

\subsubsection{Symmetry line}
\label{sec:sym}

In \fig{fss}(d), the variation of $\ebx$ with $\al$ as obtained in our 
simulations using \eq{eq:cx} is shown. Coexistence of the bound and the unbound 
phases occurs along this symmetry line, $\eb=\ebx(\al)$, which implies that the 
Hamiltonian is invariant under \ahum{swapping} the two coexisting phases. For 
the discrete Hamiltonian \eq{eq:sim} this corresponds to the operation
\begin{equation}
 (\delta_i,b_i) \mapsto (-\delta_i,1-b_i),
\end{equation}
where $\delta_i\equiv h_i-\mean{h}$ is the height deviation at lattice site~$i$ 
around the mean membrane height $\mean{h}$. A straightforward calculation shows 
that, in order for $\Ha_N$ to be invariant, we are left with the condition
\begin{equation}
 (2\mean{h}+\al)\frac{1}{N}\sum_{i=1}^{N}\delta_i =
 (\al\mean{h}-\ebx)(1-2\phi).
\end{equation}
Hence, the mean height along the symmetry line obeys $\mean{h}=-\al/2$. 
Comparing this result with an alternative calculation in 
appendix~\ref{sec:height} we find that $\mean{\phi}=1/2$ along the symmetry 
line, as expected. Moreover, from the second condition we obtain the symmetry 
line: $\ebx=-\al^2/2$. As \fig{fss}(d) shows, the agreement with the simulation 
result is excellent.

In the effective model \eq{eq:G}, spin-reversal symmetry corresponds to the 
operation $b_i \mapsto 1-b_i$. The symmetry line is now determined through
\begin{equation}
 \mu_0^\text{cx} = -\frac{1}{2}\sum_{j\neq i}\eps_{ij} = \ebx + \eps_0/2,
\end{equation}
where the sum runs over all sites $j$ excluding $i$. Plugging in the definition 
\eq{eq:eps} for the coupling energy $\eps_0$, the binding energy at coexistence 
is $\ebx=-c\al^2$ with $c=(\rgh^2/2)(1+4\eA+4\eB+\dots)$. The prefactor~$c$ 
apparently depends on the correlation length $\cor$ determining the interaction 
range, but should nevertheless converge to $c=1/2$. We have checked for $\cor=1$ 
that this is indeed the case [\fig{fss}(d)].

At the coexistence binding energy, $\ebx=-\al^2/2$, the order parameter
probability $P(\phi)$ is symmetric about $\phi=1/2$ [\fig{lt}(a)]. The latter
reflects the spin-reversal symmetry of the Ising model, which persists in the
membrane model as becomes evident from the theoretical mapping. We emphasize
that this applies to the membrane without pinning sites: in the presence of
pinning sites, spin-reversal symmetry is generally broken. Instead of using
the density of closed bonds $\phi$ as the order parameter, we could also have
used the membrane height per site $h = (1/L^2) \sum_i h_i$~\cite{weik02} since
the latter is directly coupled to the density of closed bonds. This can also
be seen in \fig{lt}(c), where we show the same snapshot as in (b), but this
time color-coded according to the membrane height. Furthermore, in a canonical
(fixed~$\phi$) simulation, and using the symmetry value $\phi=1/2$, the
binding energy always assumes the coexistence value $\ebx$ (for the Ising
model with conserved order parameter~\cite{newman.barkema:1999}, the analogue
of this condition is that, at zero magnetization, the external field is
zero). In a grand-canonical simulation, \ie where $\phi$ is allowed to
fluctuate, and using the coexistence binding energy $\ebx$, the probability
distribution of the membrane height $P(h)$ will thus be symmetric about
$h=-\al/2$.

\subsubsection{Line tension}

\begin{figure}
\centering
\includegraphics[width=\columnwidth]{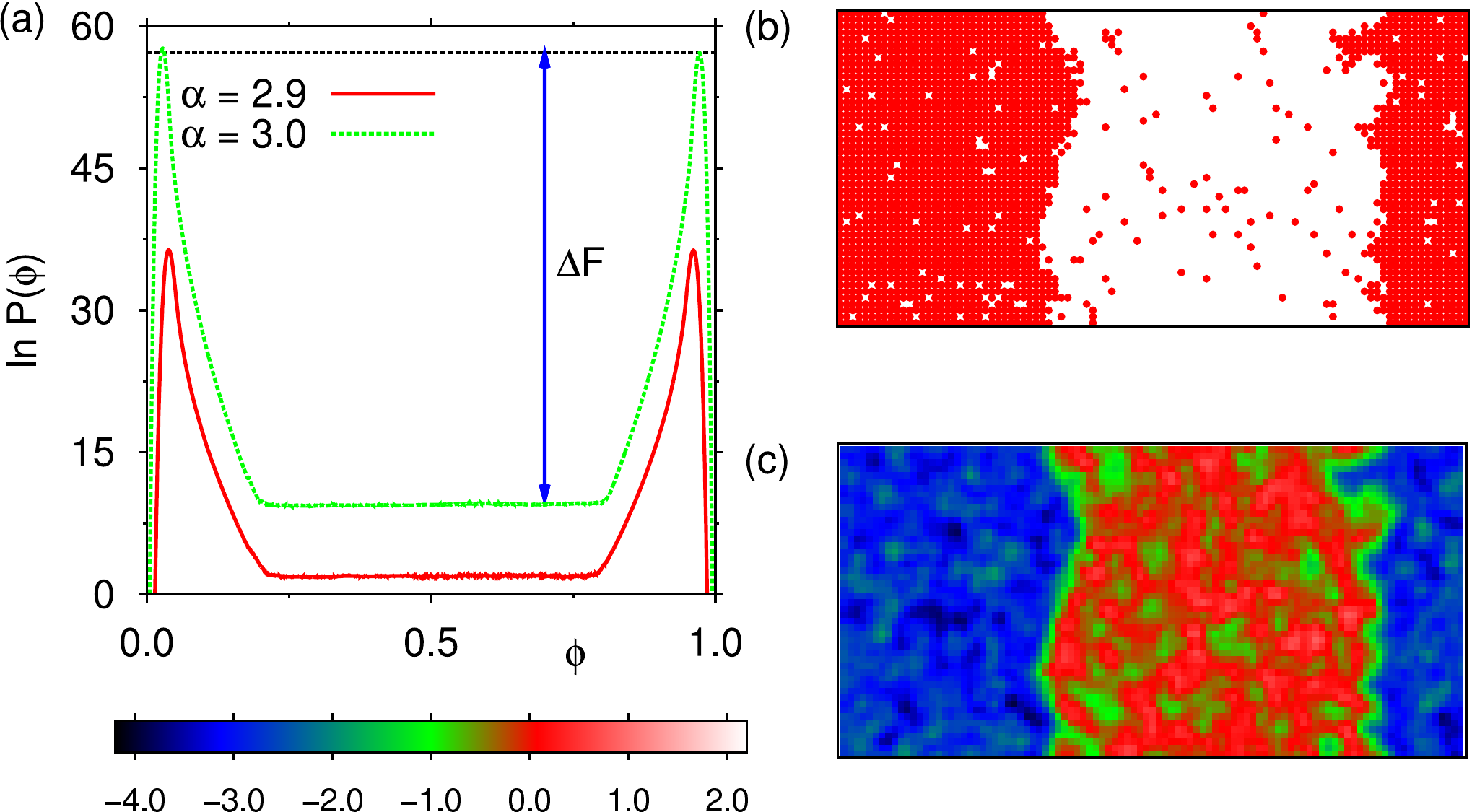}
\caption{\label{lt} (Color online) (a) Free energy $\ln P(\phi)$ at coexistence 
for two values $\al \gg \al^\ast$ using a rectangular $50 \times 100$ simulation 
lattice. The vertical double-arrow indicates the free energy cost $\Delta F$ of 
interface formation. Note the pronounced flat region between the peaks. (b) 
Snapshot of the membrane at $\phi=1/2$ and $\al=2.9$, where the dots indicate 
closed bonds. (c) The same snapshot, but color-coded according to the membrane 
height. The snapshots (b,c) clearly indicate that, in regions with a high 
density of closed bonds, the membrane is \ahum{pulled down}.}
\end{figure}

Next, we consider $\al \gg \al^\ast$, where the transition is strongly
first-order. In \fig{lt}(a), we show $\ln P(\phi)$ at coexistence for two
values of $\al$, using a rectangular $50 \times 100$ simulation box (note that
$\ln P(\phi)$ may be regarded as {\it minus} the free energy of the
system). We observe that $\ln P(\phi)$ is profoundly bimodal, which indicates
two-phase coexistence. The left peak corresponds to the unbound phase (low
density of closed bonds), the right peak to the bound phase (high density of
closed bonds), while in the region \ahum{between the peaks} both phases appear
simultaneously.  The latter follows strikingly from snapshots, see
\fig{lt}(b), which was taken at $\phi=0.5$. For this value of $\phi$, the
phases arrange in two slabs, with two interfaces perpendicular to the longest
edge, as this minimizes the total length of the line interface (whose length
then equals $L_{\rm tot}=2 L_{\rm short}$, with $L_{\rm short}=50$ the
shortest edge of the simulation box).  Provided the interfaces do not
interact, there is a region over which $\phi$ can be varied without any free
energy cost. This, apparently, is the case here, as the distributions $\ln
P(\phi)$ are essentially flat in their center regions.  Following
Binder~\cite{binder:1982}, we may then relate the free energy barrier $\Delta
F$, indicated in \fig{lt}(a) for $\al=3.0$, to the line tension $\sigma =
\Delta F / L_{\rm tot}$. Applying this equation to the distributions of
\fig{lt}(a), we obtain $\sig \simeq 0.34 \, (0.48)$ for $\al=2.9 \, (3.0)$, in
units of $k_BT$ per lattice spacing. As expected, $\sigma$ decreases upon
lowering $\al$.

\subsubsection{Summary}

For the membrane without pinning, our simulation results are thus fully 
consistent with the theoretical prediction that such a system should map onto 
the 2D Ising lattice gas. By means of finite-size scaling, 2D Ising critical 
exponents are confirmed, the quadratic variation of the coexistence binding 
energy with $\al$ is recovered, and we observe the analogue of spin-reversal 
symmetry. Furthermore, our estimate of $\al^\ast$ is in good agreement with 
\olcite{weik02}, providing an important consistency check for the composite MC 
moves proposed in this work. Finally, for $\al \gg \al^\ast$, we observe a 
first-order phase transition between bound and unbound phases, with an 
associated coexistence region where macroscopic domain formation occurs.

\subsection{The pinned membrane}

We now consider a membrane with a finite concentration of pinning sites 
$\rho>0$. We continue to use $\kappa=\gamma=1$ in \eq{eq:sim}, for which the 
case without pinning was just described (\sect{nopin}). In what follows, we 
restrict ourselves to $\rho \ll 1$, \ie the limit of low pinning concentration. 
We believe this to be the biologically most relevant situation, as well as the 
physically most interesting one (in the opposite limit $\rho \to 1$, the pinning 
sites would completely freeze the membrane, thereby trivially preventing any 
membrane-mediated phenomena from taking place).

\subsubsection{Numerical evidence for random-field disorder}

The key prediction of the theoretical mapping, \eq{eq:G}, is that, in the 
presence of pinning sites, the chemical potential becomes a quenched random 
variable (\ie dependent on the spatial location in the sample, but without any 
time dependence). We thus expect \ahum{special regions} in the membrane where 
bonds prefer to close, which are those regions where the local chemical 
potential excess is positive. To test whether such regions can be found, we 
perform canonical MC simulations at a fixed fraction of closed bonds $\phi=1/2$. 
For each lattice site~$i$, we measure the thermally averaged bond occupation 
variable $\mean{b_i}$, and use these to compute the spatial 
fluctuation~\cite{citeulike:8864903, citeulike:10877119}
\begin{equation}\label{eq:chis}
 \chi_S^2 = [\mean{b}^2] - [\mean{b}]^2, \qquad
 [\mean{b}^n] = \frac{1}{N} \sum_{i=1}^{N} \mean{b_i}^n,
\end{equation}
where the sum is over all lattice sites. In the absence of spatial preference, 
$\mean{b_i}=1/2$ for all sites, implying that $\chi_S=0$. However, if sites in 
certain regions prefer closed bonds, then $\mean{b_i}$ will be distinctly 
different from $1/2$, and consequently $\chi_S>0$.

\begin{figure}
\centering
\includegraphics[width=\columnwidth]{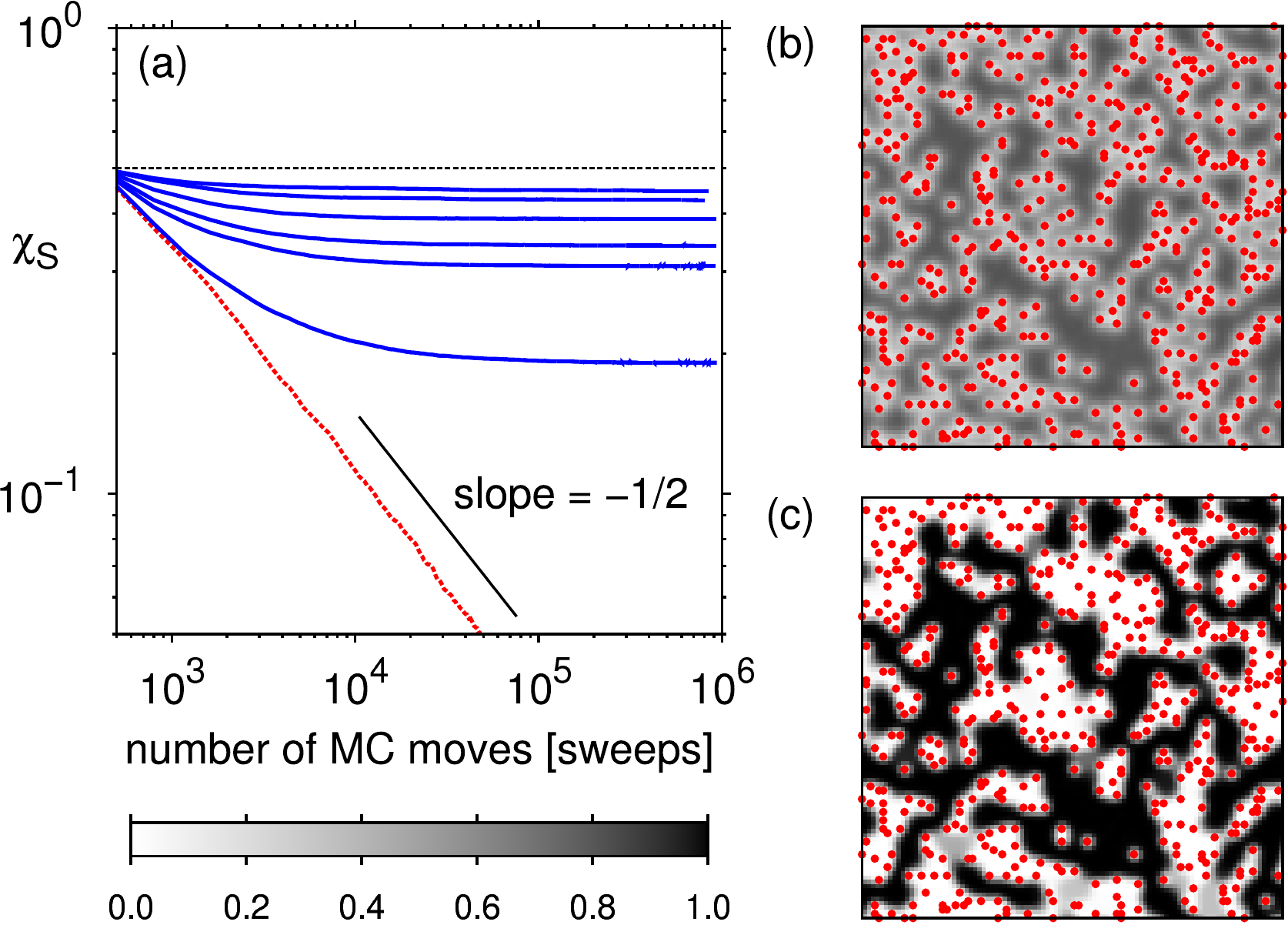}
\caption{\label{sp_pref} (Color online) (a) The \ahum{running average} of 
$\chi_S$, \eq{eq:chis}, which measures the spatial fluctuation in the thermally 
averaged bond occupation variables $\mean{b_i}$. The solid curves (blue) 
correspond to a pinned membrane for $\al=2.0, 2.5, \al^\ast, 3.0, 3.5, 4.0$ 
(from bottom to top) where $\al^\ast$ is the critical point of the unpinned 
membrane. As $\al$ increases, $\chi_S \to 1/2$ (horizontal line). The dashed 
curve (red) shows $\chi_S$ for the membrane without pinning at $\al=2.5$, in 
which case $\chi_S$ decays to zero. (b,c) \ahum{Color-maps} of the thermally 
averaged bond occupation variables $\mean{b_i}$ for $\al=2.0$ (b), and $\al=4.0$ 
(c). In both cases, the same pinning configuration (red dots) is used. (All data 
in this figure refer to $L=100$; results for the pinned membrane use $\rho=0.05$ 
and $\hp=0$).}
\end{figure}

In \fig{sp_pref}(a), we show \ahum{running averages} of $\chi_S$ as a function 
of the number of MC moves~$\tau$. For the unpinned membrane, $\chi_S$ decays to 
zero, since here there is no spatial preference. In the absence of spatial 
preference, $\chi_S$ reflects the statistical error of our simulation, and 
therefore decays $\propto \tau^{-1/2}$. In contrast, for the membrane with 
pinning sites, $\chi_S$ saturates to a finite plateau value. Note that this 
happens for all values of $\al$ considered, including those values below the 
critical point $\al^\ast$ of the unpinned membrane. By increasing $\al$, the 
plateau value increases, which means that the spatial preference becomes 
stronger. When $\al$ is small, thermal fluctuations still permit 
\ahum{excursions} of bonds into regions where the local chemical potential is 
unfavorable. In this case, $\mean{b_i}$ does not deviate much from $1/2$. As 
$\al$ increases, these thermal fluctuations are \ahum{frozen out}. In the limit 
$\al \to \infty$, the bond occupation variables are set by the groundstate of 
\eq{eq:sim} (that is: for a given configuration of pinning sites, the variables 
$b_i$ are determined by energy minimization). In this limit, $\mean{b_i}$ is 
either 0 or 1, and the fluctuation $\chi_S=1/2$. As can be seen in 
\fig{sp_pref}(a), this limiting value is indeed approached as $\al$ increases.

However, already for $\al$ much smaller, the groundstate reveals itself. To 
demonstrate this, we show \ahum{color-maps} of the thermally averaged values 
$\mean{b_i}$~\cite{citeulike:8864903, citeulike:10877119}, for two values of 
$\al$, but using the same configuration of pinning sites with $\hp=0$ 
[\fig{sp_pref}(b,c)]. Clearly visible is that the adhesion domains (dark 
regions) prefer the same locations for both values of $\al$. The only difference 
is that, with increasing $\al$, the preference becomes stronger. The structure 
of \fig{sp_pref}(c) is already close to the groundstate, since the majority of 
$\mean{b_i}$ are already close to either 0 or 1. But also for the smaller value 
of $\al$, the groundstate is visible, albeit somewhat \ahum{blurred}. Note that 
$\al$ in \fig{sp_pref}(b) is considerably below $\al^\ast$ of the unpinned 
membrane, \ie the presence of the groundstate persists deep into the 
high-temperature region. 

Our finding that, in the presence of pinning sites, $\mean{b_i}$ generally 
deviates from $1/2$ may be conceived as a local chemical potential excess 
(field) at site~$i$ given by
\begin{equation}\label{eq:chem}
 \Delta\mu_i = \ln\frac{\mean{b_i}}{1-\mean{b_i}},
\end{equation}
where the canonical ensemble with $\phi=1/2$ is assumed. Furthermore, the
\ahum{color-maps} of \fig{sp_pref}(b,c) suggest that $\Delta\mu_i$ is a
spatially random variable. Since the chemical potential is isomorphic to an
external field in the Ising model, this indeed corresponds to {\it
  random-field} disorder. Consequently, the pinned membrane belongs to the
universality class of the two-dimensional random-field Ising model. This
implies that the \ahum{freezing out} of the thermal fluctuations with
increasing $\al$ in \fig{sp_pref} is a gradual process, \ie there is no phase
transition associated with it (and nothing special happens at the critical
point $\al^\ast$ of unpinned membrane). As is well known, the random-field
Ising model in $d=2$ dimensions does not support any phase
transition~\cite{nattermann:1998}.

\begin{figure}
\centering
\includegraphics[width=0.8\columnwidth]{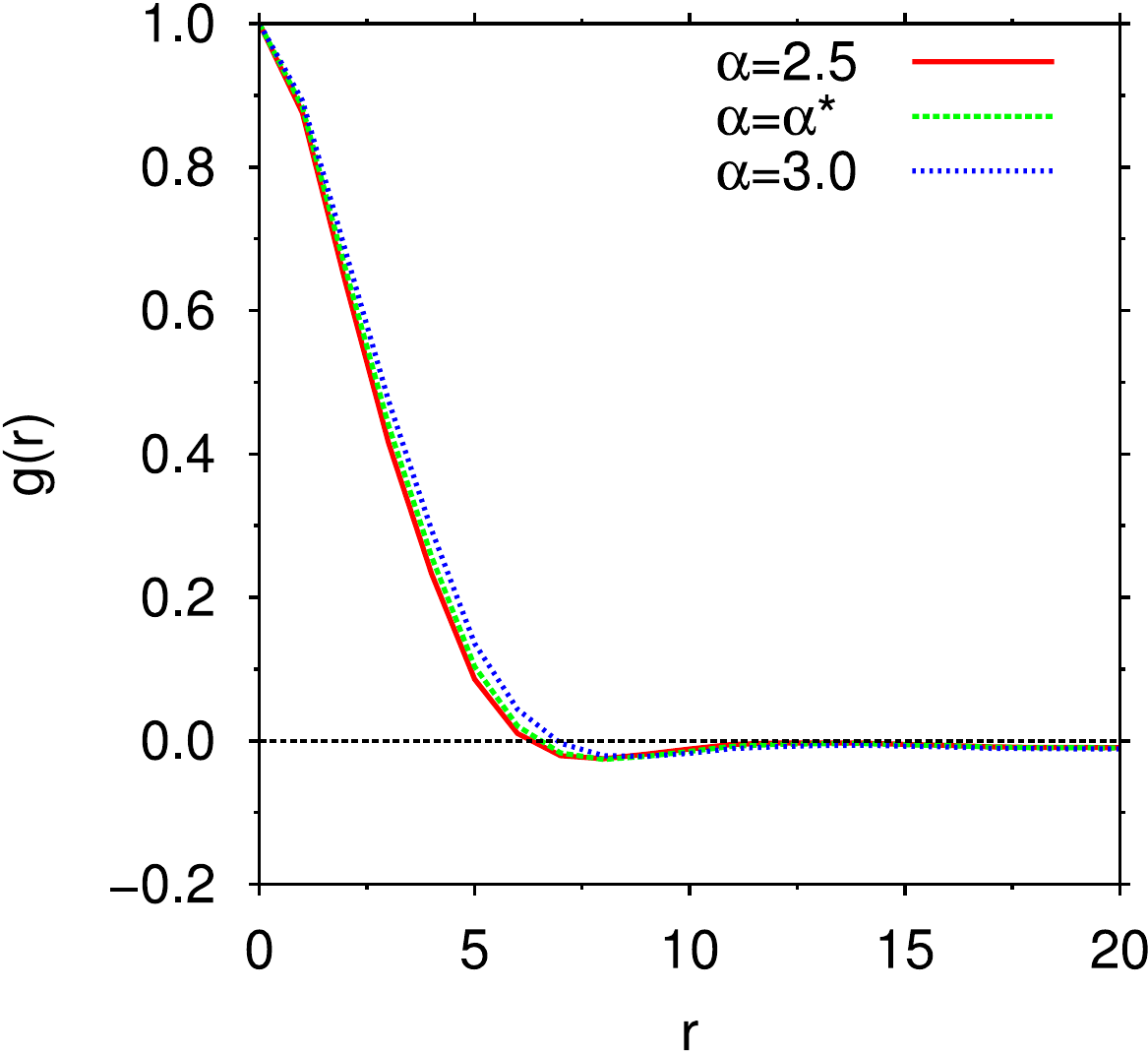}
\caption{\label{gr} (Color online) The (normalized) correlation function $g(r)$ 
of the local chemical potential excess $\Delta\mu_i$, for several values of 
$\al$. The correlations quickly decay to zero, which confirms the prediction of 
the mapping that $\Delta\mu_i$ is a spatially random variable (results refer to 
the pinned membrane with $\rho=0.05$, $\hp=0$, and $L=100$).}
\end{figure}

To confirm that $\Delta\mu_i$ truly is a spatially random variable, we have
measured the (circularly averaged) correlation function $g(r) \propto
[\Delta\mu_i \Delta\mu_j]'$, where $[\cdot]'$ denotes a spatial average over
pairs of sites $(i,j)$ a distance $r$ apart. As can be seen in \fig{gr},
$g(r)$ quickly decays to zero. For randomly distributed pinning sites, the
spatial correlations in the local chemical potential excess are thus
short-ranged. Note also that $g(r)$ depends only weakly on~$\al$. This is
consistent with the \ahum{color-maps} of \fig{sp_pref}(b,c), whose overall
topology (\ie the shape of the regions) is also remarkably insensitive
to~$\al$. The insensitivity to $\al$ shows that the {\it sign} of the local
chemical potential excess in \eq{eq:chem} is determined exclusively by the
properties of the membrane and the pinning sites (most notably the positions
of the latter, and the pinning height $\hp$). The picture that one should have
in mind, therefore, is that of ligand-receptor bonds \ahum{diffusing} through
a chemical potential landscape, but the landscape itself is quenched, \ie the
bonds do not modify nor shape it. Incidentally, for $\hp=0$,
\fig{sp_pref}(b,c) shows that the local chemical potential excess is such that
the pinning sites ``repel'' closed bonds. By using a sufficiently negative
pinning height, the reverse situation can be realized also.

\subsubsection{Adhesion domains are finite: Imry-Ma argument}

For the pinned membrane, Figs.~\ref{sp_pref} and \ref{gr} clearly show 
that the local chemical potential excess is a quenched random variable, thereby 
confirming the theoretical prediction. This rigorously rules out the formation 
of large (macroscopic) adhesion domains, by virtue of the Imry-Ma 
argument~\cite{imry75, imbrie:1984, bricmont.kupiainen:1987, 
aizenman.wehr:1989}. To see this, consider a cluster of closed bonds of linear 
size~$l$, thereby containing $\propto l^d$ bonds, and imagine to insert the 
cluster at some location in the sample. The average chemical potential excess 
(per site) over the cluster is zero, but with fluctuations that decay conform 
the central limit theorem
\begin{equation}
 \frac{1}{l^d} \sum_{i \in \rm cluster} \Delta\mu_i = 0 \pm C/l^{d/2},
\end{equation}
with $\Delta\mu_i$ given by \eq{eq:chem}, and $C$ a constant. Hence, there exist 
\ahum{preferred regions} where the cost of insertion is reduced by an amount 
$\propto l^{d/2}$. In $d=2$ dimensions (but not in $d=3$~\cite{nattermann:1998, 
citeulike:10004339}), this is sufficient to compensate the line tension, which 
scales $\sim l^{d-1}$. In the presence of pinning sites, adhesion domains thus 
no longer strive to minimize the length of their line interface. Instead, they 
seek out those regions in the sample where the local chemical potential excess 
is most favorable, precisely what is observed in \fig{sp_pref}(b,c). We thus 
obtain a stable multi-domain structure. Note the sharp contrast to the case 
without pinning, where adhesion domains do minimize their interface, thereby 
growing macroscopically large [\fig{lt}].

\subsubsection{Domain size statistics}

Having argued that adhesion domains in the presence of pinning sites remain
finite, we now present a more quantitative analysis of their size. As the
domain structure is essentially fixed by the pinning sites, and much less by
thermal effects, we restrict ourselves to $\al=3.5$, and vary only (i) the
pinning concentration $\rho$, and (ii) the lateral correlation length $\cor$
of the membrane fluctuations. The simulations in this section are again
performed in the canonical ensemble ($\phi=1/2$), and pinning height $\hp=0$
(pinning sites thus ``repel'' closed bonds). The remaining fixed parameters
are $\gamma=1$ for the non-specific potential, and system size $L=100$.

For a given sample $i=1,\ldots,K$ of pinning sites, we generate a series 
$t=1,\ldots,T$ of equilibrated snapshots. For each snapshot, we compute the 
typical domain size $R_{t,i} = 2\pi \int S_{t,i} (k) \, dk / \int k S_{t,i} (k) 
\, dk$~\cite{newman.barkema:1999}, where $S_{t,i}(k)$ is the (circularly 
averaged) static structure factor; disorder $[\cdot]$ and thermal $\mean{\cdot}$ 
averages are then computed as
\begin{equation}\label{eq:avg}
 [\mean{R^n}^m] = \frac{1}{K} \sum_{i=1}^K \left(
 \frac{1}{T} \sum_{t=1}^T R_{t,i}^n \right)^m.
\end{equation}
We will primarily be concerned with the average domain size $[\mean{R}]$, the
thermal fluctuations $\chi_T^2 \equiv \mean{R^2} - \mean{R}^2$, and the
disorder fluctuations $\chi_D^2 \equiv [\mean{R}^2] - [\mean{R}]^2$.

\begin{figure}
\centering
\includegraphics[width=\columnwidth]{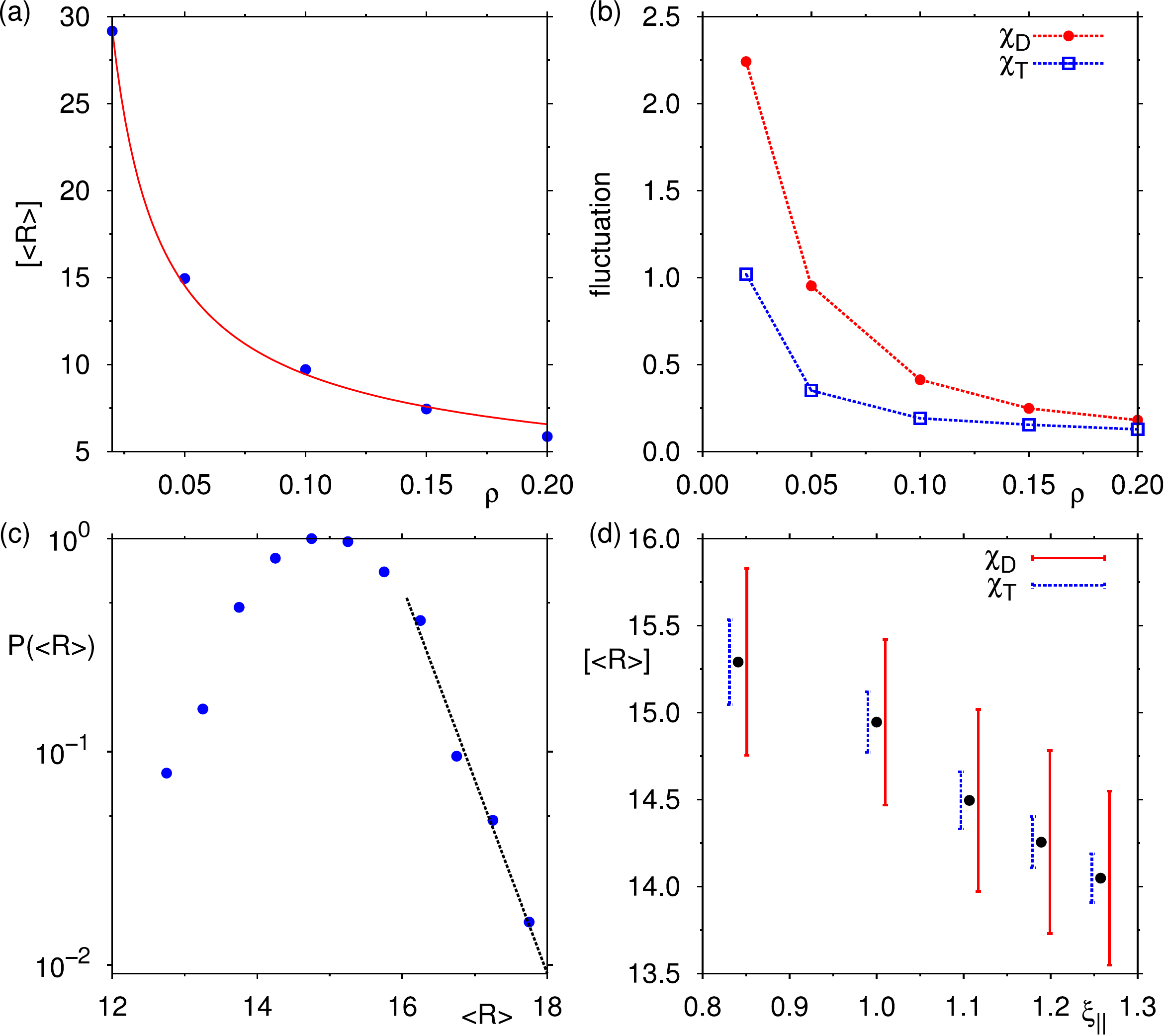}
\caption{\label{fig:dom} (Color online) (a) Average domain size $[\mean{R}]$ as 
function of pinning density $\rho$ (symbols are simulation data; the curve is a 
fit to \eq{eq:rho}). (b) The disorder fluctuation $\chi_D$ and the thermal 
fluctuation $\chi_T$ in the domain size as function of $\rho$. (c) Distribution 
of the thermally averaged domain size between samples for $\rho=0.05$. Note the 
logarithmic vertical scale. The dashed line shows the result of an exponential 
fit (see details in text). (d) Variation of the typical domain size $[\mean{R}]$ 
with the lateral correlation length $\cor$ (dots). Also shown for each 
measurement are the magnitudes of the disorder $\chi_D$ and thermal fluctuations 
$\chi_T$. (Note: All averages in this figure were computed using \eq{eq:avg} 
with $T \sim 500$, $K=20$ (a,b,d), and $K=500$ (c). The data in (a-c) were 
obtained for $\kappa=1$, while in (d) $\kappa$ was varied.)}
\end{figure}

In \fig{fig:dom}(a), we plot the average domain size $[\mean{R}]$ as function of 
the pinning concentration $\rho$. As might be expected, the domain size 
decreases with increasing concentration. Following theoretical predictions for 
the random-field Ising model~\cite{binder:1983, citeulike:2842269, 
citeulike:10788402}, we anticipate that
\begin{equation}\label{eq:rho}
 \ln \, [\mean{R}] \propto \rho^{-p} \quad,
\end{equation}
with exponent $p>0$, whereby we assume that $[\mean{R}]$ is the analogue
of the \ahum{break-up} length. For bimodal and Gaussian random-fields
$p=2$, but since it is {\it a priori} unclear how pinning sites compare
to such fields, we leave $p$ as a free parameter. The curve in
\fig{fig:dom}(a) shows the corresponding fit of \eq{eq:rho} to our data, where
$p \simeq 0.25$ was used, and the agreement is quite reasonable.

Next, we consider the magnitude of the thermal $\chi_T$ and disorder 
fluctuations $\chi_D$ [\fig{fig:dom}(b)]. As $\rho$ increases, the fluctuations 
become smaller, but we always find that $\chi_D > \chi_T$. The disorder 
fluctuations thus dominate, as we had already announced previously. In 
\fig{fig:dom}(c), we plot the distribution $P(\mean{R})$ of the thermally 
averaged domain size $\mean{R}$ between samples for $\rho=0.05$ (the width of 
this distribution thus reflects the disorder fluctuation $\chi_D$). Experiments 
have indicated that the tail of the distribution is exponential~\cite{gov06}. 
The dashed line in \fig{fig:dom}(c) shows a fit to the tail of the simulated 
distribution using an exponential function of the form $P(x) \propto e^{-bx}$. 
The fit captures the data rather well, where $b \simeq 2.1$ was used. 

Last but not least, we show in \fig{fig:dom}(d) the variation of the average 
domain size $[\mean{R}]$ with the lateral correlation length $\cor$. As $\cor$ 
increases, there is a mild decrease of the domain size. However, on the scale of 
the (dominating) disorder fluctuations, the effect may be difficult to observe 
in experiments.

\subsubsection{The case of \ahum{neutral} pinning sites}

\begin{figure}
\centering
\includegraphics[width=\columnwidth]{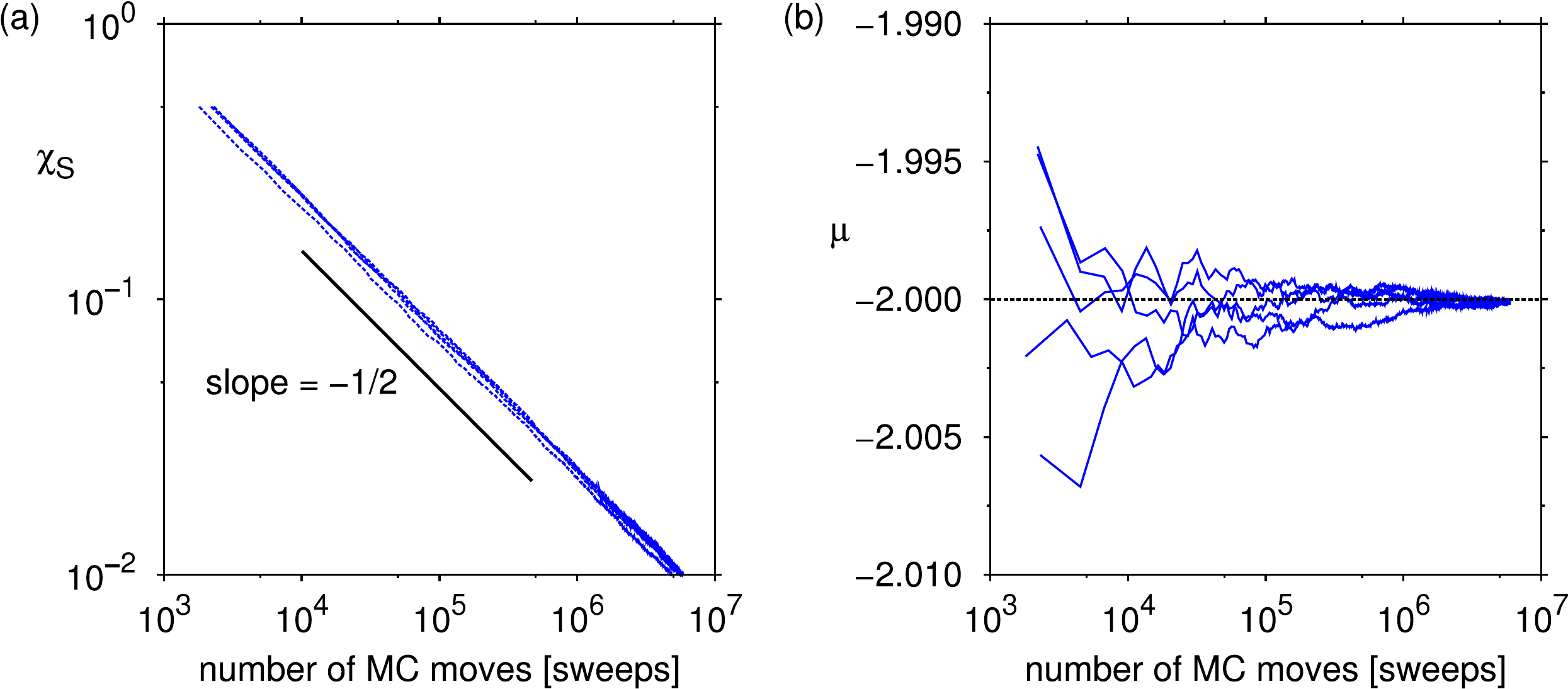}
\caption{\label{dil} (Color online) Simulation results for the pinned membrane 
using pinning height $\hp=\hp^\ast=-\al/2$. In this case, there is no 
random-field effect. (a) The \ahum{running average} of the spatial fluctuation 
$\chi_S$. Clearly visible is that $\chi_S \to 0$. (b) The \ahum{running average} 
of the chemical potential $\mu$, which approaches $\ebx$ (horizontal line). All 
data refer to $\rho=0.05, \al=2.0, L=40$. In each plot, results from five 
independent simulation runs are shown.}
\end{figure}

As can be seen in \fig{sp_pref}(b,c), for $\hp=0$ the pinning sites ``repel''
closed bonds. When the pinning height is sufficiently negative, the reverse
situation is obtained and pinning sites attract closed bonds (we have checked
this case for $\hp=-\al$). Therefore, it is plausible that for some
intermediate height $\hp^\ast$ the pinning sites become neutral. This special
height is given by the symmetry height of the unpinned membrane:
$\hp^\ast=-\al/2$ (\sect{sec:sym}). Consequently, when $\hp=\hp^\ast$, we no
longer expect the local chemical potential excess to be a quenched random
variable, but rather $\Delta\mu_i=0$ for all sites.

To test this assertion, we performed a canonical simulation ($\phi=1/2$) using
pinning concentration $\rho=0.05$, pinning height $\hp=\hp^\ast$, and
$\al=2.0$.  In \fig{dil}(a), we plot the \ahum{running average} of the spatial
fluctuation $\chi_S$ in the thermally averaged bond occupation variables
(\eq{eq:chis}). The figure strikingly shows $\chi_S \to 0$, which confirms
that the local chemical potential excess has indeed vanished. As for the
unpinned membrane, $\chi_S$ now reflects the statistical error of our MC
simulation, and therefore decays $\propto \tau^{-1/2}$, $\tau$ being the
number of MC moves. This result is to be contrasted with \fig{sp_pref}(a),
where $\chi_S$ for the pinned membrane converged to a finite value. Hence,
even though the pinning concentration is the same in both cases, choosing
$\hp=\hp^\ast$ completely destroys the random-field effect. In \fig{dil}(b),
we show the \ahum{running average} of the chemical potential obtained via
Widom insertion~\cite{citeulike:3404574, frenkel.smit:2001}. Interestingly,
$\mu$ converges exactly onto $\ebx=-\al^2/2$ of the unpinned
membrane. Apparently, for neutral pinning sites, the symmetry line of the
unpinned membrane is restored.

At $\hp=\hp^\ast$, the Imry-Ma argument thus no longer applies. We therefore 
expect a critical point again, at $\al=\al'$, and macroscopic adhesion domains 
for $\al>\al'$. The value of $\al'$ should increase with $\rho$, and $\lim_{\rho 
\to 0} \al'=\al^\ast$ of the unpinned membrane. The quenched disorder induced by 
neutral pinning sites is of the {\it dilution} type. The diluted Ising model is 
similar to the standard Ising model, but with a (small) fraction of randomly 
selected sites removed from the lattice~\cite{citeulike:4197235}. Note that this 
type of quenched disorder does not break spin reversal symmetry, consistent with 
our observation that $\mu \to \ebx$ [\fig{dil}(b)].

In realistic situations, we do not expect the pinning sites to be neutral, nor 
that the pinning height is the same for all pinning sites (as assumed in the 
present study). In all these cases, it is the random-field scenario that 
applies. Nevertheless, for our fundamental understanding we felt inclined to 
discuss the neutral case also.

\section{Summary and Conclusions}

We have studied a minimal model~\cite{weik02,weik07} describing the specific
adhesion of a cell (or biomimetic vesicle) to another cell, the extracellular
matrix, or a substrate. The model incorporates the arguably most important
mechanism: the coupling of thermal fluctuations of the cell membrane to the
state of the receptor-ligand pairs responsible for the cell attachment. The
latter are effectively described as either open or closed bonds. Integrating
out the membrane fluctuations this model can be mapped exactly onto the
two-dimensional lattice gas (or Ising model). This mapping confirms previous
numerical evidence of a phase separation~\cite{weik02} between an unbound and
a bound state. For fixed concentration of mobile bonds this scenario implies
the formation of a single macroscopic domain of bonds. Moreover, we have
exactly determined the critical point, and we have demonstrated numerically
that the full model indeed belongs to the Ising universality class as
anticipated from the theoretical mapping.

A striking observation in experiments is the absence of a single large
domain. Rather, adhesion domains of finite size form. This observation is
often explained as either caused by active processes, or the ``corralling'' of
adhesion proteins in compartments~\cite{kusu05}. This putative hindrance of
protein diffusion is a purely dynamical approach, which does not alter
equilibrium properties. Quite in contrast, here we have demonstrated that
already in equilibrium finite sized adhesion domains are implied in a wide
variety of contexts if one takes into account \ahum{pinning sites} that
locally suppress membrane height fluctuations. These pinning sites model
perturbations that necessarily occur \emph{in vivo} due to the crowded, highly
non-ideal composition of cell membranes and the anchoring of the cytoskeleton
to the ECM or other cells.

We have presented both analytical calculations and numerical evidence that the
presence of pinning sites corresponds to quenched disorder which, in the
language of the Ising model, induces a random field that prevents macroscopic
domain formation in two dimensions~\cite{imry75, imbrie:1984,
  bricmont.kupiainen:1987, aizenman.wehr:1989}. This is in sharp contrast to
the type of quenched disorder where the positions of ligand-receptor bonds are
random. The latter constitutes a marginal perturbation~\cite{lipo96} and does
not fundamentally alter the scenario of~\fig{lt}, \ie macroscopic domain
formation above some critical~$\al$. In contrast, random-field disorder is a
relevant perturbation.  It requires that the disorder couples linearly to the
order parameter. We have demonstrated here that this condition is typically
fulfilled for membrane adhesion, where membrane height pinning (disorder)
couples to bond formation (order parameter $\phi$). Hence, the fundamentals of
statistical physics can be used to explain adhesion domains of finite size in
equilibrium.

\acknowledgments

We acknowledge financial support by the Alexander-von-Humboldt foundation (TS)
and the Deutsche Forschungsgemeinschaft (RV: Emmy Noether VI~483).

\appendix
\section{Height correlations}
\label{sec:corr}

Using periodic boundaries the separation profile $h(\x)$ is expanded into
Fourier modes through
\begin{equation*}
  h(\x) = \sum_{\q} h_\q e^{-\im\q\cdot\x}, \qquad
  h_\q = \frac{1}{A} \IInt{^2\x}{A}{} h(\x) e^{\im\q\cdot\x}
\end{equation*}
with $ h_\q=h'_\q+\im h''_\q$. Since the separation field is real
\begin{equation*}
  h(\x) = h_0 + 2\sum_{\q\in\Q} [h'_\q\cos(\q\cdot\x) +
  h''_\q\sin(\q\cdot\x)],
\end{equation*}
where the set $\Q$ contains the independent wave vectors excluding $q=0$.

For a free membrane in the absence of both bonds and pinning sites the
Hamiltonian Eq.~\eqref{eq:H} in Fourier space reads
\begin{equation*}
  \beta\Ha_0 = \frac{1}{2}\gam_0h_0^2+\frac{1}{2}\sum_{\q\in\Q} 
  \gam_q(h_\q'^2+h_\q''^2)
\end{equation*}
with $\gam_0\equiv\beta A\gam$ and $\gam_q\equiv 2\gam_0[1+(q\cor)^4]$. Hence,
the mean height is $\mean{h(\x)}_0=0$ and the height correlations read
\begin{equation*}
  m(|\x-\x'|) \equiv \mean{h(\x)h(\x')}_0
  = \frac{1}{\gam_0} + 4\sum_{\q\in\Q}\frac{\cos\q\cdot(\x-\x')}{\gam_q}.
\end{equation*}
The height correlations between two points on the membrane decay on the length
scale $\cor$. For large separations they reach
$m(r\ra\infty)=\mean{h_0^2}_0=1/\gam_0$ due to the zero mode fluctuations. For
large $A$ we can neglect this contribution. Replacing the sum over discrete
wave vectors by an integral we then obtain
\begin{equation*}
  m(r) \approx \frac{\cor^4}{2\pi\beta\kap}
  \IInt{q}{0}{\infty}\frac{qJ_0(qr)}{1+(q\cor)^4},
\end{equation*}
where $J_0$ is the zero-order Bessel function of the first kind. Performing
the final integration leads to Eq.~\eqref{eq:m} for the height correlations of
a free membrane.

\section{Effective free energy}
\label{sec:free}

The full partition sum of the system reads
\begin{multline}
  \label{eq:Z}
  \mathcal Z = \sum_{\{b_i\}} \Path{h(\x)} \Int{h_1\cdots\dd h_N}
  e^{-\beta\Ha_N} \\ \times
  \prod_{i=1}^N\delta(h(\x_i)-h_i) \prod_{k=N+1}^{M}\ell\delta(h(\x_k)-\hp),
\end{multline}
where the Hamiltonian $\Ha_N$ is given in Eq.~\eqref{eq:HN}. The $M=N+n$ total
constraints due to the $N$ bonds and the presence of $n$ pinning sites are
represented through $\delta$-functions, where $\ell$ is a microscopic
length. We normalize the functional measure such that $\mathcal Z=1$ for a
free membrane.

As usual, we represent the $\delta$-functions as integrals
\begin{equation*}
  \delta(h(\x_i)-h_i) = \frac{1}{2\pi}\Int{\lam_i}
  e^{\im\lam_i[h(\x_i)-h_i]},
\end{equation*}
where we complement $h_i=\hp$ for $i>N$. This allows us to write the partition
sum Eq.~\eqref{eq:Z} involving a single exponential
\begin{widetext}
\begin{multline*}
  \mathcal Z = \sum_{\{b_i\}} \Path{h(\x)} \frac{\ell^n}{(2\pi)^M} 
  \Int{\lam_1\cdots\dd\lam_M} \Int{h_1\cdots\dd h_N} 
  \exp\left\{ -\beta\Ha_0 + c_0h_0 + \sum_{\q\in\Q}(c_\q'h_\q'+c_\q''h_\q'')
    -\im\sum_{i=1}^M \lam_ih_i + \beta\eb\sum_{i=1}^N b_i \right\}
\end{multline*}
\end{widetext}
with coefficients
\begin{gather*}
  c_0 \equiv \im\sum_{i=1}^M \lam_i - \beta\al\sum_{i=1}^Nb_i, \\
  c_\q' \equiv 2\im\sum_{i=1}^M\lam_i\cos\q\cdot\x_i - 2\beta\al\sum_{i=1}^N
  b_i\cos\q\cdot\x_i, \\
  c_\q'' \equiv 2\im\sum_{i=1}^M\lam_i\sin\q\cdot\x_i - 2\beta\al\sum_{i=1}^N
  b_i\sin\q\cdot\x_i.
\end{gather*}
Note that the first sum runs over all sites whereas the second sum only
includes the sites where bonds are present. Performing the Gaussian
integrations over the height modes $\{h_\q\}$ and the auxiliary variables
$\{\lam_i\}$ we obtain
\begin{multline}
  \label{eq:Z:fin}
  \mathcal Z = \frac{\ell^n}{(2\pi)^{M/2}}\sqrt{\det\mat m^{-1}}
  \sum_{\{b_i\}} \Int{h_1\cdots\dd h_N} \\ \times
  \exp\left\{ -\frac{1}{2}\sum_{ij}^M (\mat m^{-1})_{ij}h_jh_j -
    \beta\sum_{i=1}^N b_i[\al h_i - \eb] \right\}.
\end{multline}
The prefactor is independent of the state of the bonds. While in principle it
depends on the quenched disorder of the pinning sites it does not contribute
to thermal averages. Setting $\hp=0$ the terms with indices $i>N$
corresponding to the pinning sites drop out of the sum and we obtain
Eq.~\eqref{eq:G:int}.

\section{Mean height}
\label{sec:height}

The derivation presented in the previous section is easily modified to
calculate a generating function, from which arbitrary moments can be
obtained. For example, replacing $c_0\ra c_0+\lam$ we obtain the generating
function $\mathcal Z(\lam)$ from which the mean height follows as
\begin{equation}
  \mean{h_0} = \left.\pd{\ln\mathcal Z(\lam)}{\lam}\right|_{\lam=0}.
\end{equation}
Repeating the calculation for $n=0$ we obtain
\begin{equation}
  \mean{h_0} = -\frac{\al N}{\gam A}\mean{\phi}.
\end{equation}



\begin{thebibliography}{42}
\expandafter\ifx\csname natexlab\endcsname\relax\def\natexlab#1{#1}\fi
\expandafter\ifx\csname bibnamefont\endcsname\relax
  \def\bibnamefont#1{#1}\fi
\expandafter\ifx\csname bibfnamefont\endcsname\relax
  \def\bibfnamefont#1{#1}\fi
\expandafter\ifx\csname citenamefont\endcsname\relax
  \def\citenamefont#1{#1}\fi
\expandafter\ifx\csname url\endcsname\relax
  \def\url#1{\texttt{#1}}\fi
\expandafter\ifx\csname urlprefix\endcsname\relax\def\urlprefix{URL }\fi
\providecommand{\bibinfo}[2]{#2}
\providecommand{\eprint}[2][]{\url{#2}}

\bibitem[{\citenamefont{Geiger et~al.}(2001)\citenamefont{Geiger, Bershadsky,
  Pankov, and Yamada}}]{geig01}
\bibinfo{author}{\bibfnamefont{B.}~\bibnamefont{Geiger}},
  \bibinfo{author}{\bibfnamefont{A.}~\bibnamefont{Bershadsky}},
  \bibinfo{author}{\bibfnamefont{R.}~\bibnamefont{Pankov}}, \bibnamefont{and}
  \bibinfo{author}{\bibfnamefont{K.~M.} \bibnamefont{Yamada}},
  \bibinfo{journal}{Nat. Rev. Mol. Cell Biol.} \textbf{\bibinfo{volume}{2}},
  \bibinfo{pages}{793} (\bibinfo{year}{2001}).

\bibitem[{\citenamefont{Parsons et~al.}(2010)\citenamefont{Parsons, Horwitz,
  and Schwartz}}]{pars10}
\bibinfo{author}{\bibfnamefont{J.~T.} \bibnamefont{Parsons}},
  \bibinfo{author}{\bibfnamefont{A.~R.} \bibnamefont{Horwitz}},
  \bibnamefont{and} \bibinfo{author}{\bibfnamefont{M.~A.}
  \bibnamefont{Schwartz}}, \bibinfo{journal}{Nat. Rev. Mol. Cell Biol.}
  \textbf{\bibinfo{volume}{11}}, \bibinfo{pages}{633} (\bibinfo{year}{2010}).

\bibitem[{\citenamefont{Nicolas et~al.}(2004)\citenamefont{Nicolas, Geiger, and
  Safran}}]{nico04}
\bibinfo{author}{\bibfnamefont{A.}~\bibnamefont{Nicolas}},
  \bibinfo{author}{\bibfnamefont{B.}~\bibnamefont{Geiger}}, \bibnamefont{and}
  \bibinfo{author}{\bibfnamefont{S.~A.} \bibnamefont{Safran}},
  \bibinfo{journal}{Proc. Natl. Acad. Sci. U.S.A.}
  \textbf{\bibinfo{volume}{101}}, \bibinfo{pages}{12520}
  (\bibinfo{year}{2004}).

\bibitem[{\citenamefont{Gov}(2006)}]{gov06}
\bibinfo{author}{\bibfnamefont{N.~S.} \bibnamefont{Gov}},
  \bibinfo{journal}{Biophys. J.} \textbf{\bibinfo{volume}{91}},
  \bibinfo{pages}{2844} (\bibinfo{year}{2006}).

\bibitem[{\citenamefont{Bell}(1978)}]{bell78}
\bibinfo{author}{\bibfnamefont{G.~I.} \bibnamefont{Bell}},
  \bibinfo{journal}{Science} \textbf{\bibinfo{volume}{200}},
  \bibinfo{pages}{618} (\bibinfo{year}{1978}).

\bibitem[{\citenamefont{Weikl et~al.}(2002)\citenamefont{Weikl, Andelman,
  Komura, and Lipowsky}}]{weik02}
\bibinfo{author}{\bibfnamefont{T.}~\bibnamefont{Weikl}},
  \bibinfo{author}{\bibfnamefont{D.}~\bibnamefont{Andelman}},
  \bibinfo{author}{\bibfnamefont{S.}~\bibnamefont{Komura}}, \bibnamefont{and}
  \bibinfo{author}{\bibfnamefont{R.}~\bibnamefont{Lipowsky}},
  \bibinfo{journal}{Eur. Phys. J. E} \textbf{\bibinfo{volume}{8}},
  \bibinfo{pages}{59} (\bibinfo{year}{2002}).

\bibitem[{\citenamefont{Krobath et~al.}(2007)\citenamefont{Krobath, Sch\"utz,
  Lipowsky, and Weikl}}]{krob07}
\bibinfo{author}{\bibfnamefont{H.}~\bibnamefont{Krobath}},
  \bibinfo{author}{\bibfnamefont{G.~J.} \bibnamefont{Sch\"utz}},
  \bibinfo{author}{\bibfnamefont{R.}~\bibnamefont{Lipowsky}}, \bibnamefont{and}
  \bibinfo{author}{\bibfnamefont{T.~R.} \bibnamefont{Weikl}},
  \bibinfo{journal}{EPL} \textbf{\bibinfo{volume}{78}}, \bibinfo{pages}{38003}
  (\bibinfo{year}{2007}).

\bibitem[{\citenamefont{Speck et~al.}(2010)\citenamefont{Speck, Reister, and
  Seifert}}]{spec10b}
\bibinfo{author}{\bibfnamefont{T.}~\bibnamefont{Speck}},
  \bibinfo{author}{\bibfnamefont{E.}~\bibnamefont{Reister}}, \bibnamefont{and}
  \bibinfo{author}{\bibfnamefont{U.}~\bibnamefont{Seifert}},
  \bibinfo{journal}{Phys. Rev. E} \textbf{\bibinfo{volume}{82}},
  \bibinfo{pages}{021923} (\bibinfo{year}{2010}).

\bibitem[{\citenamefont{Weil and Farago}(2010)}]{weil10}
\bibinfo{author}{\bibfnamefont{N.}~\bibnamefont{Weil}} \bibnamefont{and}
  \bibinfo{author}{\bibfnamefont{O.}~\bibnamefont{Farago}},
  \bibinfo{journal}{Eur. Phys. J. E} \textbf{\bibinfo{volume}{33}},
  \bibinfo{pages}{81} (\bibinfo{year}{2010}).

\bibitem[{\citenamefont{Farago}(2011)}]{fara11}
\bibinfo{author}{\bibfnamefont{O.}~\bibnamefont{Farago}},
  \emph{\bibinfo{title}{Advances in Planar Lipid Bilayers and Liposomes}}
  (\bibinfo{publisher}{Elsevier}, \bibinfo{year}{2011}),
  vol.~\bibinfo{volume}{14}, chap.~\bibinfo{chapter}{5}, pp.
  \bibinfo{pages}{129--155}, ISBN \bibinfo{isbn}{9780123877208}.

\bibitem[{\citenamefont{Lipowsky}(1996)}]{lipo96}
\bibinfo{author}{\bibfnamefont{R.}~\bibnamefont{Lipowsky}},
  \bibinfo{journal}{Phys. Rev. Lett.} \textbf{\bibinfo{volume}{77}},
  \bibinfo{pages}{1652} (\bibinfo{year}{1996}).

\bibitem[{\citenamefont{Zhang and Wang}(2008)}]{zhan08}
\bibinfo{author}{\bibfnamefont{C.-Z.} \bibnamefont{Zhang}} \bibnamefont{and}
  \bibinfo{author}{\bibfnamefont{Z.-G.} \bibnamefont{Wang}},
  \bibinfo{journal}{Phys. Rev. E} \textbf{\bibinfo{volume}{77}},
  \bibinfo{pages}{021906} (\bibinfo{year}{2008}).

\bibitem[{\citenamefont{Atilgan and Ovryn}(2009)}]{atil09}
\bibinfo{author}{\bibfnamefont{E.}~\bibnamefont{Atilgan}} \bibnamefont{and}
  \bibinfo{author}{\bibfnamefont{B.}~\bibnamefont{Ovryn}},
  \bibinfo{journal}{Biophys. J.} \textbf{\bibinfo{volume}{96}},
  \bibinfo{pages}{3555} (\bibinfo{year}{2009}).

\bibitem[{\citenamefont{Tanaka and Sackmann}(2005)}]{tana05}
\bibinfo{author}{\bibfnamefont{M.}~\bibnamefont{Tanaka}} \bibnamefont{and}
  \bibinfo{author}{\bibfnamefont{E.}~\bibnamefont{Sackmann}},
  \bibinfo{journal}{Nature} \textbf{\bibinfo{volume}{437}},
  \bibinfo{pages}{656} (\bibinfo{year}{2005}).

\bibitem[{\citenamefont{Mossman and Groves}(2007)}]{moss07}
\bibinfo{author}{\bibfnamefont{K.}~\bibnamefont{Mossman}} \bibnamefont{and}
  \bibinfo{author}{\bibfnamefont{J.~T.} \bibnamefont{Groves}},
  \bibinfo{journal}{Chem.~Soc.~Rev.} \textbf{\bibinfo{volume}{36}},
  \bibinfo{pages}{46} (\bibinfo{year}{2007}).

\bibitem[{\citenamefont{Bruinsma et~al.}(2000)\citenamefont{Bruinsma, Behrisch,
  and Sackmann}}]{brui00}
\bibinfo{author}{\bibfnamefont{R.}~\bibnamefont{Bruinsma}},
  \bibinfo{author}{\bibfnamefont{A.}~\bibnamefont{Behrisch}}, \bibnamefont{and}
  \bibinfo{author}{\bibfnamefont{E.}~\bibnamefont{Sackmann}},
  \bibinfo{journal}{Phys. Rev. E} \textbf{\bibinfo{volume}{61}},
  \bibinfo{pages}{4253} (\bibinfo{year}{2000}).

\bibitem[{\citenamefont{Cuvelier and Nassoy}(2004)}]{cuve04}
\bibinfo{author}{\bibfnamefont{D.}~\bibnamefont{Cuvelier}} \bibnamefont{and}
  \bibinfo{author}{\bibfnamefont{P.}~\bibnamefont{Nassoy}},
  \bibinfo{journal}{Phys. Rev. Lett.} \textbf{\bibinfo{volume}{93}},
  \bibinfo{pages}{228101} (\bibinfo{year}{2004}).

\bibitem[{\citenamefont{Reister-Gottfried
  et~al.}(2008)\citenamefont{Reister-Gottfried, Sengupta, Lorz, Sackmann,
  Seifert, and Smith}}]{reis08}
\bibinfo{author}{\bibfnamefont{E.}~\bibnamefont{Reister-Gottfried}},
  \bibinfo{author}{\bibfnamefont{K.}~\bibnamefont{Sengupta}},
  \bibinfo{author}{\bibfnamefont{B.}~\bibnamefont{Lorz}},
  \bibinfo{author}{\bibfnamefont{E.}~\bibnamefont{Sackmann}},
  \bibinfo{author}{\bibfnamefont{U.}~\bibnamefont{Seifert}}, \bibnamefont{and}
  \bibinfo{author}{\bibfnamefont{A.-S.} \bibnamefont{Smith}},
  \bibinfo{journal}{Phys. Rev. Lett.} \textbf{\bibinfo{volume}{101}},
  \bibinfo{pages}{208103} (\bibinfo{year}{2008}).

\bibitem[{\citenamefont{Limozin and Sengupta}(2009)}]{limo09}
\bibinfo{author}{\bibfnamefont{L.}~\bibnamefont{Limozin}} \bibnamefont{and}
  \bibinfo{author}{\bibfnamefont{K.}~\bibnamefont{Sengupta}},
  \bibinfo{journal}{ChemPhysChem} \textbf{\bibinfo{volume}{10}},
  \bibinfo{pages}{2752} (\bibinfo{year}{2009}).

\bibitem[{\citenamefont{Smith and Sackmann}(2009)}]{smit09}
\bibinfo{author}{\bibfnamefont{A.-S.} \bibnamefont{Smith}} \bibnamefont{and}
  \bibinfo{author}{\bibfnamefont{E.}~\bibnamefont{Sackmann}},
  \bibinfo{journal}{ChemPhysChem} \textbf{\bibinfo{volume}{10}},
  \bibinfo{pages}{66} (\bibinfo{year}{2009}).

\bibitem[{\citenamefont{Kusumi et~al.}(2005)\citenamefont{Kusumi, Nakada,
  Ritchie, Murase, Suzuki, Murakoshi, Kasai, Kondo, and Fujiwara}}]{kusu05}
\bibinfo{author}{\bibfnamefont{A.}~\bibnamefont{Kusumi}},
  \bibinfo{author}{\bibfnamefont{C.}~\bibnamefont{Nakada}},
  \bibinfo{author}{\bibfnamefont{K.}~\bibnamefont{Ritchie}},
  \bibinfo{author}{\bibfnamefont{K.}~\bibnamefont{Murase}},
  \bibinfo{author}{\bibfnamefont{K.}~\bibnamefont{Suzuki}},
  \bibinfo{author}{\bibfnamefont{H.}~\bibnamefont{Murakoshi}},
  \bibinfo{author}{\bibfnamefont{R.~S.} \bibnamefont{Kasai}},
  \bibinfo{author}{\bibfnamefont{J.}~\bibnamefont{Kondo}}, \bibnamefont{and}
  \bibinfo{author}{\bibfnamefont{T.}~\bibnamefont{Fujiwara}},
  \bibinfo{journal}{Annu. Rev. Biophys. Biomol. Struct}
  \textbf{\bibinfo{volume}{34}}, \bibinfo{pages}{351} (\bibinfo{year}{2005}).

\bibitem[{\citenamefont{Imry and Ma}(1975)}]{imry75}
\bibinfo{author}{\bibfnamefont{Y.}~\bibnamefont{Imry}} \bibnamefont{and}
  \bibinfo{author}{\bibfnamefont{S.-k.} \bibnamefont{Ma}},
  \bibinfo{journal}{Phys. Rev. Lett.} \textbf{\bibinfo{volume}{35}},
  \bibinfo{pages}{1399} (\bibinfo{year}{1975}).

\bibitem[{\citenamefont{Imbrie}(1984)}]{imbrie:1984}
\bibinfo{author}{\bibfnamefont{J.~Z.} \bibnamefont{Imbrie}},
  \bibinfo{journal}{Phys.~Rev. Lett.} \textbf{\bibinfo{volume}{53}},
  \bibinfo{pages}{1747} (\bibinfo{year}{1984}).

\bibitem[{\citenamefont{Bricmont and
  Kupiainen}(1987)}]{bricmont.kupiainen:1987}
\bibinfo{author}{\bibfnamefont{J.}~\bibnamefont{Bricmont}} \bibnamefont{and}
  \bibinfo{author}{\bibfnamefont{A.}~\bibnamefont{Kupiainen}},
  \bibinfo{journal}{Phys.~Rev. Lett.} \textbf{\bibinfo{volume}{59}},
  \bibinfo{pages}{1829} (\bibinfo{year}{1987}).

\bibitem[{\citenamefont{Aizenman and Wehr}(1989)}]{aizenman.wehr:1989}
\bibinfo{author}{\bibfnamefont{M.}~\bibnamefont{Aizenman}} \bibnamefont{and}
  \bibinfo{author}{\bibfnamefont{J.}~\bibnamefont{Wehr}},
  \bibinfo{journal}{Phys.~Rev. Lett.} \textbf{\bibinfo{volume}{62}},
  \bibinfo{pages}{2503} (\bibinfo{year}{1989}).

\bibitem[{\citenamefont{Weikl and Lipowsky}(2007)}]{weik07}
\bibinfo{author}{\bibfnamefont{T.}~\bibnamefont{Weikl}} \bibnamefont{and}
  \bibinfo{author}{\bibfnamefont{R.}~\bibnamefont{Lipowsky}},
  \emph{\bibinfo{title}{Advances in Planar Lipid Bilayers and Liposomes}}
  (\bibinfo{publisher}{Elsevier}, \bibinfo{year}{2007}),
  vol.~\bibinfo{volume}{5}, chap.~\bibinfo{chapter}{4}, pp.
  \bibinfo{pages}{64--127}.

\bibitem[{\citenamefont{Helfrich}(1978)}]{helf78}
\bibinfo{author}{\bibfnamefont{W.}~\bibnamefont{Helfrich}},
  \bibinfo{journal}{Z.~Naturforsch.} \textbf{\bibinfo{volume}{33a}},
  \bibinfo{pages}{305} (\bibinfo{year}{1978}).

\bibitem[{\citenamefont{Li and Kardar}(1991)}]{li91}
\bibinfo{author}{\bibfnamefont{H.}~\bibnamefont{Li}} \bibnamefont{and}
  \bibinfo{author}{\bibfnamefont{M.}~\bibnamefont{Kardar}},
  \bibinfo{journal}{Phys. Rev. Lett.} \textbf{\bibinfo{volume}{67}},
  \bibinfo{pages}{3275} (\bibinfo{year}{1991}).

\bibitem[{\citenamefont{Virnau and M\"{u}ller}(2004)}]{virnau.muller:2004}
\bibinfo{author}{\bibfnamefont{P.}~\bibnamefont{Virnau}} \bibnamefont{and}
  \bibinfo{author}{\bibfnamefont{M.}~\bibnamefont{M\"{u}ller}},
  \bibinfo{journal}{J.~Chem.~Phys.} \textbf{\bibinfo{volume}{120}},
  \bibinfo{pages}{10925} (\bibinfo{year}{2004}).

\bibitem[{\citenamefont{Ferrenberg and
  Swendsen}(1988)}]{ferrenberg.swendsen:1988}
\bibinfo{author}{\bibfnamefont{A.~M.} \bibnamefont{Ferrenberg}}
  \bibnamefont{and} \bibinfo{author}{\bibfnamefont{R.~H.}
  \bibnamefont{Swendsen}}, \bibinfo{journal}{Phys.~Rev. Lett.}
  \textbf{\bibinfo{volume}{61}}, \bibinfo{pages}{2635} (\bibinfo{year}{1988}).

\bibitem[{\citenamefont{Fischer and Vink}(2011)}]{citeulike:8864903}
\bibinfo{author}{\bibfnamefont{T.}~\bibnamefont{Fischer}} \bibnamefont{and}
  \bibinfo{author}{\bibfnamefont{R.~L.~C.} \bibnamefont{Vink}},
  \bibinfo{journal}{J.~Chem.~Phys.} \textbf{\bibinfo{volume}{134}},
  \bibinfo{pages}{055106} (\bibinfo{year}{2011}).

\bibitem[{\citenamefont{Newman and Barkema}(1999)}]{newman.barkema:1999}
\bibinfo{author}{\bibfnamefont{M.~E.~J.} \bibnamefont{Newman}}
  \bibnamefont{and} \bibinfo{author}{\bibfnamefont{G.~T.}
  \bibnamefont{Barkema}}, \emph{\bibinfo{title}{Monte Carlo Methods in
  Statistical Physics}} (\bibinfo{publisher}{Clarendon Press},
  \bibinfo{address}{Oxford}, \bibinfo{year}{1999}).

\bibitem[{\citenamefont{Binder}(1982)}]{binder:1982}
\bibinfo{author}{\bibfnamefont{K.}~\bibnamefont{Binder}},
  \bibinfo{journal}{Phys. Rev. A} \textbf{\bibinfo{volume}{25}},
  \bibinfo{pages}{1699} (\bibinfo{year}{1982}).

\bibitem[{\citenamefont{Fischer et~al.}(2012)\citenamefont{Fischer,
  Jelger~Risselada, and Vink}}]{citeulike:10877119}
\bibinfo{author}{\bibfnamefont{T.}~\bibnamefont{Fischer}},
  \bibinfo{author}{\bibfnamefont{H.}~\bibnamefont{Jelger~Risselada}},
  \bibnamefont{and} \bibinfo{author}{\bibfnamefont{R.~L.~C.}
  \bibnamefont{Vink}}, \bibinfo{journal}{Phys. Chem. Chem. Phys.}
  (\bibinfo{year}{2012}).

\bibitem[{\citenamefont{Nattermann}(1998)}]{nattermann:1998}
\bibinfo{author}{\bibfnamefont{T.}~\bibnamefont{Nattermann}}, in
  \emph{\bibinfo{booktitle}{Spin Glasses and Random Fields}}, edited by
  \bibinfo{editor}{\bibfnamefont{A.~P.} \bibnamefont{Young}}
  (\bibinfo{publisher}{World Scientific}, \bibinfo{address}{Singapore},
  \bibinfo{year}{1998}), p. \bibinfo{pages}{277}.

\bibitem[{\citenamefont{Vink et~al.}(2010)\citenamefont{Vink, Fischer, and
  Binder}}]{citeulike:10004339}
\bibinfo{author}{\bibfnamefont{R.~L.~C.} \bibnamefont{Vink}},
  \bibinfo{author}{\bibfnamefont{T.}~\bibnamefont{Fischer}}, \bibnamefont{and}
  \bibinfo{author}{\bibfnamefont{K.}~\bibnamefont{Binder}},
  \bibinfo{journal}{Phys.~Rev. E} \textbf{\bibinfo{volume}{82}},
  \bibinfo{pages}{051134} (\bibinfo{year}{2010}).

\bibitem[{\citenamefont{Binder}(1983)}]{binder:1983}
\bibinfo{author}{\bibfnamefont{K.}~\bibnamefont{Binder}}, \bibinfo{journal}{Z.
  Phys. B} \textbf{\bibinfo{volume}{50}}, \bibinfo{pages}{343}
  (\bibinfo{year}{1983}).

\bibitem[{\citenamefont{Grinstein and Ma}(1983)}]{citeulike:2842269}
\bibinfo{author}{\bibfnamefont{G.}~\bibnamefont{Grinstein}} \bibnamefont{and}
  \bibinfo{author}{\bibfnamefont{S.~K.} \bibnamefont{Ma}},
  \bibinfo{journal}{Phys.~Rev. B} \textbf{\bibinfo{volume}{28}},
  \bibinfo{pages}{2588} (\bibinfo{year}{1983}).

\bibitem[{\citenamefont{Sepp\"{a}l\"{a}
  et~al.}(1998)\citenamefont{Sepp\"{a}l\"{a}, Pet\"{a}j\"{a}, and
  Alava}}]{citeulike:10788402}
\bibinfo{author}{\bibfnamefont{E.~T.} \bibnamefont{Sepp\"{a}l\"{a}}},
  \bibinfo{author}{\bibfnamefont{V.}~\bibnamefont{Pet\"{a}j\"{a}}},
  \bibnamefont{and} \bibinfo{author}{\bibfnamefont{M.~J.} \bibnamefont{Alava}},
  \bibinfo{journal}{Phys.~Rev. E} \textbf{\bibinfo{volume}{58}},
  \bibinfo{pages}{R5217} (\bibinfo{year}{1998}).

\bibitem[{\citenamefont{Widom}(1963)}]{citeulike:3404574}
\bibinfo{author}{\bibfnamefont{B.}~\bibnamefont{Widom}},
  \bibinfo{journal}{J.~Chem.~Phys.} \textbf{\bibinfo{volume}{39}},
  \bibinfo{pages}{2808} (\bibinfo{year}{1963}).

\bibitem[{\citenamefont{Frenkel and Smit}(2001)}]{frenkel.smit:2001}
\bibinfo{author}{\bibfnamefont{D.}~\bibnamefont{Frenkel}} \bibnamefont{and}
  \bibinfo{author}{\bibfnamefont{B.}~\bibnamefont{Smit}},
  \emph{\bibinfo{title}{Understanding Molecular Simulation}}
  (\bibinfo{publisher}{Academic Press}, \bibinfo{address}{San Diego},
  \bibinfo{year}{2001}).

\bibitem[{\citenamefont{Mazzeo and K\"{u}hn}(1999)}]{citeulike:4197235}
\bibinfo{author}{\bibfnamefont{G.}~\bibnamefont{Mazzeo}} \bibnamefont{and}
  \bibinfo{author}{\bibfnamefont{R.}~\bibnamefont{K\"{u}hn}},
  \bibinfo{journal}{Phys.~Rev. E} \textbf{\bibinfo{volume}{60}},
  \bibinfo{pages}{3823} (\bibinfo{year}{1999}).
\end{thebibliography}
\end{document}